\documentclass[superscriptaddress,aps,prb,english,floatfix,twocolumn,10pt]{revtex4-2}
\usepackage[T1]{fontenc}
\usepackage[utf8]{inputenc}
\usepackage[english]{babel}
\usepackage{graphicx}
\usepackage{float}
\usepackage{natbib}
\usepackage{amssymb}
\usepackage{amsmath}
\usepackage{mathtools}
\usepackage{braket}
\usepackage{subfigure}
\usepackage{pgfplots}
\usepackage{csquotes}
\usepackage{hhline}
\usepackage{amssymb}
\usepackage{xr}

\usepackage{dcolumn}
\usepackage{tabularx}
\usepackage[colorlinks=true,linkcolor=blue,citecolor=blue,urlcolor=blue]{hyperref}
\usepackage{longtable}
\usepackage{listings}
\usepackage{color}
\usepackage[normalem]{ulem}
\usepackage[export]{adjustbox}

\DeclareMathOperator{\Tr}{Tr}
\DeclareMathOperator{\diag}{diag}

\begin{document}

\preprint{APS/123-QED}

\title{\textbf{Monitored free fermions under periodic driving}}
\author{Aditi Chakrabarty}
\author{Alexander D. Mirlin}
\author{Igor Poboiko}
\affiliation{\mbox{Institute for Quantum Materials and Technologies, Karlsruhe Institute of Technology, 76131 Karlsruhe, Germany}}
\affiliation{\mbox{Institut f\"ur Theorie der Kondensierten Materie, Karlsruhe Institute of Technology, 76131 Karlsruhe, Germany}}
\date{\today}

\begin{abstract}
We investigate analytically and numerically a one-dimensional periodically driven free-fermionic system subjected to monitoring of the local particle density. Based on the analytical approach that describes the long-wavelength physics of the time-dependent Hamiltonian in the field-theoretical language using the nonlinear sigma-model (NLSM), we reveal that driving does not alter the universality class of the problem. As a consequence, the system retains the area-law behavior in the thermodynamic limit, with an intermediate diffusive regime giving rise to logarithmic growth of entanglement entropy for a small monitoring rate. At the same time, driving leads to a renormalization of the bare coupling constant of the NLSM, 
which controls the space-time ``conductivity'' in the diffusive regime. We derive the analytic form of this renormalization, which becomes particularly strong in the case of a ``maximally symmetric'' drive and sufficiently short driving period. In addition, we employ the Wiener-Hopf method to investigate the ballistic-diffusive crossover.
These analytical predictions are corroborated by numerical simulations of the von-Neumann entanglement entropy and the density correlation function.
Our numerical results clearly demonstrate that, with an increase in the system size, there are successive crossovers from ballistic to diffusive behavior and ultimately to localization. Furthermore, in the diffusive regime, we observe weak-localization corrections that are in agreement with the analytical predictions of the NLSM. Overall, our results provide a unified analytical and numerical framework for understanding the effects of monitoring in time-modulated fermionic systems, paving a way for broader investigations of driven quantum matter.
\end{abstract}

\maketitle

\section{Introduction}\label{Sec:Introduction}
In an isolated quantum system, the time evolution of a pure state is unitary, as determined by the Schr\"odinger equation, which generically drives the system towards a highly entangled state. On the other hand, in monitored setups
local measurements tend to disentangle the wavefunction by projecting it onto  eigenstates of the measurement operator depending on the outcome. The competition between the measurements and the unitary dynamics may drive a macroscopic phase transition, which is known as a measurement-induced phase transition (MIPT) and occurs at a critical measurement rate, $\gamma_c$. For $\gamma<\gamma_c$, the steady state exhibits an entanglement entropy that grows with the subsystem size, whereas for $\gamma > \gamma_c$, the ``quantum Zeno'' phase is realized that exhibits an area-law behavior \cite{Li_2018, Li_2019}. In recent years, the MIPTs arising from both random projective measurements \cite{Smith_2019,Skinner_2019,Jian_2020,Bao_2020,Gullans_2020,Fidkowski_2021,Agrawal_2022} and continuous measurements \cite{Cao_2019, Szyniszewski_2019, Szyniszewski_2020, Alberton_2021,Turkeshi_2021} have attracted much attention. The case of continuous monitoring 
effectively corresponds to the limiting case in which the system is subjected to very frequent, yet weak, measurements, as can be realized experimentally within platforms based on superconducting qubits \cite{Katz_2006,Vijay_2012,Huard_2016} and cold atoms \cite{Yang_2018}.

Investigations on the effects of quantum measurements on quantum correlations---and, in particular, on entanglement---represent an increasing interest to quantum-information properties of condensed-matter systems.
More generally, recent years have witnessed a renaissance in the science of quantum information in areas like quantum computation, quantum cryptography and quantum teleportation \cite{Vaidman_1994,Braunstein_1998,Furusawa_1998,Gisin_2002,Grosshans_2002,Scarani_2009}. In particular, it was discovered that the Gaussian states and Gaussian measurements (employing homodyne detection \cite{Lvovsky_2009}) comes with several advantages in quantum information processing \cite{Braunstein_2005,Eisert_2010,Weedbrook_2012}. On the experimental side, the optical components used in such circuits are also readily available. Motivated by these advances and by a fundamental challenge of understanding quantum phases and their universal properties, the investigation of MIPTs, originally carried out in the Haar-random unitary and Clifford circuits, has been extended to the Gaussian ones \cite{Cao_2019,Turkeshi_2021, Alberton_2021, Buchhold_2021, Coppola_2022, Carollo_2022, Szyniszewski_2023, Jian_2023, Fava_2023, Poboiko_2023, Poboiko_2024, Chahine_2023, Lumia_2023, Starchl_2025, FavaNahum2024}. 

Importantly, these Gaussian states are realized by the many-body free-fermion systems. It was demonstrated that, for complex free fermions and spatial dimensionality $d=1$, the system is in the area-law phase in the thermodynamic limit (spatial scale $\ell \to \infty$) for any monitoring rate $\gamma >0$ \cite{Poboiko_2023, FavaNahum2024}; see also Ref.~\cite{Niederegger_2026}. 
 For  $d > 1$, a MIPT emerges between the area-law phase and a phase with  a long-range entanglement, $S(\ell) \propto \ell^{d-1} \ln \ell$ \cite{Poboiko_2024, Chahine_2023}. These results were obtained by a mapping to a replica nonlinear sigma-model (NLSM) with its subsequent renormalization-group (RG) treatment, and were also supported by numerical simulations. This formalism reveals a clear analogy  \cite{Poboiko_2023,Poboiko_2024} between the measurement-induced physics
in $d$ dimensions and the physics of Anderson localization problem in $d+1$ dimensions. Such an analogy manifests itself in close similarity between the corresponding NLSM field theories (which are, however, characterized by different symmetry classes and replica limits) and leads to remarkable parallels for physical observables. 
A related broad discussion of field theories of random non-unitary dynamics, with connections to the tenfold symmetry classification of disordered systems \cite{Zirnbauer_1996,Altland_1997} and to Anderson-localization theories, can be found in Refs.~\cite{Jian_2022,Jian_2023}. 

While a one-dimensional system of complex free fermions is in the area-law phase in the $\ell \to \infty$ limit, an MIPT
emerges if one considers a system of Majorana fermions
(thus breaking the charge conservation) \cite{Fava_2023} or by introducing inter-particle interactions \cite{Poboiko_2025,Guo_2025}. This highlights the pivotal role of symmetries in the physics of monitored systems.
It is also worth mentioning that since one can seamlessly interpolate between
continuous and projective measurements \cite{Jacobs_2006,Szyniszewski_2020}, they can be expected to give rise to the same physics from the point of view of the MIPT universality class. Indeed, this is confirmed by the fact that one can map systems with these different types of monitoring onto the same NLSM and by numerical studies.

In a parallel development, the periodic driving has emerged as a powerful route to realizing novel quantum phases and dynamical phenomena that have no counterparts in static systems, motivating extensive studies of time-dependent (Floquet) quantum systems \cite{Hanggi_1998,Stockmann_1999,Ivanov_2008,Poli_2011,Chen_2011,Bitter_2016}. The time-periodic drive has also been extensively investigated for its ability to host rich topological phases \cite{Kiagawa_2011,Kundu_2014,Karen_2020}. The outstanding success of the Floquet theory in understanding the physics of driven closed systems has led to a great effort in investigating open quantum systems subjected to periodic drive \cite{Prosen_2011,Iwahori_2016,Reimer_2018,Magazzu_2018,Scopa_2019,Ideka_2021}. However, the interplay of periodic drive and monitoring remains poorly understood. 

In this context, a natural and important question to be explored is how a time-periodic drive affects the measurement-induced physics of free-fermion systems. Recently, this problem was addressed in Ref.~\cite{Modak_2025}. The authors of that work interpreted their numerical results in terms of the Berezinskii-Kosterlitz-Thouless (BKT) transition between the area-law and critical (logarithmic) phases. 
On the analytical side, the authors supported their conclusions by a reference to the sine-Gordon model proposed in the context of free-fermion measurement-induced phenomena  in Ref.~\cite{Buchhold_2021}. If true, the BKT-transition conclusion of Ref.~\cite{Modak_2025} would imply that the periodic driving changes the universality class compared to an undriven system where no such transition takes place in one dimension \cite{Poboiko_2023}.

With the above motivations, in this work, we reconsider the problem of a monitored free-fermion system subjected to a time-periodic drive, by combining analytical and numerical approaches. On the analytical side, we derive the effective field theory (NLSM) and elucidate the role of symmetry in governing the monitored dynamics. We show that the periodic drive does not affect the form and the symmetry of the NLSM. This implies that, for any measurement rate $\gamma > 0$, the steady state exhibits an area-law entanglement in the large-$\ell$ limit, as in the absence of driving, precluding the emergence of a MIPT. 
At the same time, we show that the bare coupling constant of the theory is renormalized by the periodic drive, and this renormalization can be very strong in some range of parameters. 

We corroborate these analytical findings by numerical simulations based on the stochastic Schr\"odinger equation (SSE), where the fermionic density is continuously monitored. Our numerical results demonstrate that, with an increase in $\ell$, there are successive crossovers from ballistic to diffusive behavior and ultimately to localization. Furthermore, in the diffusive regime, we observe weak-localization (WL) corrections that are in agreement with the analytical predictions of the NLSM.
The numerical study thus provides clear support to our analytical conclusion that the periodic drive does not alter the symmetry and universality class of the phenomenon. 

The rest of the paper is structured as follows. At first, we introduce the time-periodic monitored free fermion model in Sec.~\ref{Sec:Model}. 
This is followed by the presentation of the analytical approach in Sec.~\ref{Sec:Analytic}, wherein we present the field-theoretical description of the long-wavelength behavior of the system based on the NLSM field theory (Sec.~\ref{Subsec:NLSM}), discuss analytical predictions of the NLSM theory for physical observables (Sec.~\ref{sec:analytics_observables}) and the Wiener-Hopf treatment of the ballistic-to-diffusion crossover (Sec.~\ref{Subsec:WH_integral}).
Next, we turn to numerical simulations in Sec.~\ref{Sec:Numerical_results}. In Sec.~\ref{Subsec:Entanglement_entropy}, we present our numerical results on the entanglement entropy. 
We relate these findings to the analytical predictions for the bare conductivity, the WL correction, and the saturation value of the entanglement entropy in the strong-localization regime in Secs.~\ref{Subsubsec:g_0_comparison}, \ref{Subsubsec:WL}, and \ref{Subsubsec:SL}, respectively. Further, we numerically explore the behavior of the density correlation function in Sec.~\ref{Subsec:Density_correlation}. At the end, in Sec.~\ref{Sec:Conclusions}, we summarize the key results and provide perspectives on this work. Some details of our analysis are relegated to Appendices.

\section{Microscopic model}
\label{Sec:Model}

We consider a one-dimensional chain of free spinless fermions with a time-dependent tight-binding nearest-neighbor hopping amplitude under the periodic boundary conditions. The Hamiltonian describing the unitary part of the evolution  is defined as
\begin{eqnarray}
\mathcal{H}(t)= J(t)\displaystyle\sum_i (c_{i+1}^\dagger c_i+ \text{H.c.}),
\label{Eq:Hamiltonian}
\end{eqnarray}
where $c_i^{\dagger}$ ($c_i$) denotes the fermionic creation (annihilation) operator at site $i$ of a one-dimensional lattice containing $L$ sites and $J(t)$ involves a time-periodic driving. Specifically, we choose $J(t)$ to be constant within each of the half-periods: 
\begin{equation}
\label{Eq:Jt}
J(t) = \begin{cases}
+J \,, & 0 < t \le T/2 \,, \\
-J+\epsilon \,, & T/2 < t \le T \,.
\end{cases}\end{equation}
Here, the parameter $\epsilon$ controls the asymmetry of the hopping amplitudes in the two halves of the quenched driving protocol with period $T$. Note that the Hamiltonian considered in Eq.~\eqref{Eq:Hamiltonian} commutes with the total particle number operator, as well as the translation operator.

In addition to the dynamics governed by the Hamiltonian $\mathcal{H}(t)$, the system is subjected to continuous quantum measurements of the  local fermion densities $n_i = c_i^\dagger c_i$, which make the overall evolution non-unitary. Continuous measurements act as a limiting scenario where the amount of information extracted per measurement and the duration of such measurement simultaneously approach zero \cite{Wiseman_1993_2}. Specifically, we employ the quantum state diffusion protocol \cite{Gisin_1992,Barun_2002}, where the evolution of the wavefunction is dictated by the stochastic Schr\"odinger equation (SSE) \cite{Gisin_1992, Gisin_1998, Jacobs_2006,Alonso_2017} given by 
\begin{equation}
d\ket{\Psi_{t}}=\left(-i\mathcal{H}(t)\,dt+\sum_{i}\Big(d\xi_{i,t}M_{i,t}-\frac{\gamma}{2}M^{\,2}_{i,t}\,dt\Big)\right)\ket{\Psi_{t}},
\label{Eq:SSE}
\end{equation}
where $\gamma$ controls the strength of measurements, $M_{i,t}=n_i-\langle n_i\rangle_t$, and the Gaussian white noise $d\xi_{i,t}$  satisfies $\overline{d\xi_{i,t}}=0$ and $\overline{d\xi_{i,t}\,d\xi_{j,t}}=\gamma\,dt\,\delta_{i,j}$. Let us recall that the first term comes from the usual unitary evolution of the Hamiltonian, while the second term incorporates the stochastic effects of the measurement-induced noise from the It{\^o} increments. The last term corresponds to a deterministic back action from the measurement apparatus. Such a dynamical evolution has been experimentally realized through homodyne detection in quantum optical systems \cite{Wiseman_1993}.
Details of the numerical implementation of the SSE as given in Eq.~\eqref{Eq:SSE}, are presented in Appendix~\ref{app:SSE}.

We note that the model that we study  is the same as was considered in Ref.~\cite{Modak_2025}.

\section{Analytical treatment} 
\label{Sec:Analytic}

\subsection{Field theoretical description}
\label{Subsec:NLSM}
The analytical description of the SSE \eqref{Eq:SSE} is based on the fermionic replicated Keldysh path integral description \cite{Poboiko_2023,Poboiko_2025}.
Each replica describes a \emph{linear} evolution of the non-normalized many-body state for a given quantum trajectory labeled by set of functions $\zeta_i(t)$. Note that the Born's rule probability governing the non-trivial distribution over different realizations of the measurement protocol is encoded in the replica limit $R \to 1$. 
The field theory is then formulated in terms of Grassmann fields $\psi_x(t)$, which additionally carry replica index $a = 1,\dots,R$ and Keldysh index $\tau = \pm$ for forward- and backward-time evolution, with the following Lagrangian:
\begin{equation}
\label{eq:FermionicLagrangian}
\mathcal{L}=\psi^{\dagger}\left(i\partial_{t}-{\cal H}(t)-i\zeta(t)\hat{\tau}_{z}\right)\psi.
\end{equation}
Here $\mathcal{H}(t)$ and $\zeta(t)$ are single-particle nearest-neighbor Hamiltonian and diagonal measurement-induced stochastic noise matrices, respectively, and $\hat{\tau}_z$ is the Pauli matrix acting on the Keldysh space. The measurements are described by the Stratonovich Gaussian white noise with the correlation function $\overline{\zeta_{i}(t)\zeta_{j}(t^{\prime})}=\gamma\delta_{ij}\delta(t-t^{\prime})$, with $\zeta_i(t) = d\xi_{i,t} / dt$.

\begin{figure}
    \centering
    \includegraphics[width=1.0\columnwidth]{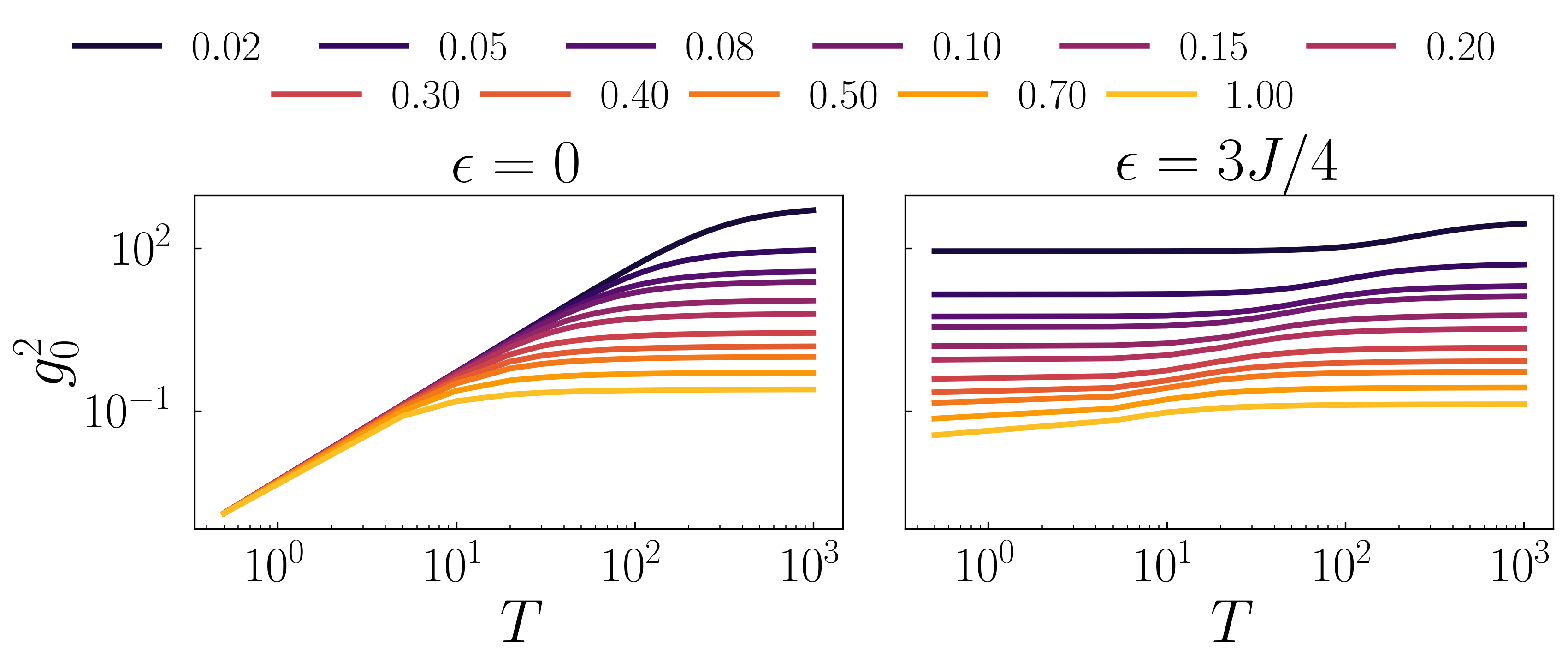}
    \caption{
    Analytical results for $g_0^2$ as obtained in Eq.~\eqref{eq:g0}, for different measurement strengths $\gamma/J$ at $\epsilon= 0$ and $\epsilon=3J/4$.
}
    \label{Fig:Fig_1}
\end{figure}

With a suitably chosen basis (specifically, replacing $\psi \mapsto \tau_x \psi$; see Ref.~\cite{Poboiko_2025} for details), the Lagrangian \eqref{eq:FermionicLagrangian} describes a $(1+1)$-dimensional chiral disordered system. 
Furthermore, the Hamiltonian given in Eq.~\eqref{Eq:Hamiltonian} has a particle-hole symmetry, since one can perform a gauge transformation which makes it purely imaginary \cite{FavaNahum2024,Poboiko_2025}. Let us emphasize that this symmetry is not affected by the time dependence of ${\cal H}(t)$ resulting from the drive.
As a consequence, from the point of view of the Altland-Zirnbauer tenfold symmetry classification of disordered systems, this specific system belongs to the Hamiltonian chiral orthogonal symmetry class (BDI).
Microscopically, this implies presence of a continuous Lie symmetry group, corresponding to simultaneous transformations of the spinors $\psi$ and $\psi^\dagger$ in the Keldysh and replica spaces \cite{Poboiko_2025}.
These massless modes, whose structure is dictated entirely by the symmetries of the model, govern the long-wavelength behavior of the system, and the field theory which describes them takes the form of the NLSM. 
Importantly, the measurement-induced ``disorder'' is local both in space and time, which leads to an effective space-time ``isotropization''. This implies that, with a suitable choice of time units, the (Euclidean) Lagrangian density acquires the following isotropic in space-time $r_{\mu}=(vt,x)$ form:
\begin{equation}
\label{eq:NLSMAction}
{\cal L}[\hat{U}]=\frac{g_0}{4}\Tr\left(\partial_{\mu}\hat{U}^{\dagger}\partial_{\mu}\hat{U}\right).
\end{equation}
The matrices $\hat{U}={\cal V}\hat{\sigma}_{y}{\cal V}^{T}\hat{\sigma}_{y}$ span the target space of the BDI NLSM manifold $\mathrm{SU}(2R)/\mathrm{USp}(2R)$, parametrized by unitary matrices ${\cal V}\in\mathrm{SU}(2R)$ acting in the auxiliary two-dimensional particle-hole and $R$-dimensional replica spaces. Here, $\hat{\sigma}_y$ is the Pauli matrix acting in the particle-hole space, and the stabilizer symplectic subgroup $\mathrm{USp}(2R)$ is determined by the constraint ${\cal V}\hat{\sigma}_{y}{\cal V}^{T}=\hat{\sigma}_{y}$ which leaves matrix $\hat{U}$ invariant.

The dimensionless coupling constant $g_0$, which enters the action in Eq.~\eqref{eq:NLSMAction}, plays the role of effective ``conductivity'', whose bare value is determined by the microscopic details of the model. It governs the relative ``strength'' of the quantum fluctuations, with $g_0 \gg 1$ corresponding to the weak-coupling regime. As we show in Appendices~\ref{app:crossover},\,\ref{app:g0}, for the case of the chosen drive as specified in Eq.~\eqref{Eq:Jt}, it is given by the following expression:
\begin{equation}
\label{eq:g0}
g_{0}=\frac{1}{2\gamma}\sqrt{\bar{J}^{2}+(\delta J)^{2}\,{\cal I}\left(\frac{\gamma T}{4}\right)},
\end{equation}
where $\bar{J} = \epsilon / 2$ is the average value of the hopping constant over the entire stroboscopic period, $\delta J = J - \epsilon / 2$  is the amplitude of the drive, and the dimensionless function ${\cal I}(p)$ is given by
${\cal I}(p)=1 - \tanh(p)/p$. The dependence of $g_0$ on the parameters of the model is illustrated in Fig.~\ref{Fig:Fig_1}. 
At large $T \gg \gamma^{-1}$, Eq.~\eqref{eq:g0} predicts a saturation of $g_0$ since ${\cal I}(p) \to 1$ at $p\to \infty$. As seen in Fig.~\ref{Fig:Fig_1}, this saturation takes place around $T\gamma \approx 10$.

The universal NLSM field theory, as given in Eq.~\eqref{eq:NLSMAction}, emerges at sufficiently large spatial scales $\ell \gg \ell_0$, with the ``mean-free path'' $\ell_0$ playing the role of the ultraviolet (UV) cutoff of the theory. 
The scale $\ell_0$ is related to the conductivity $g_0$ via $\ell_0 = 2\sqrt{2} g_0$, see Appendix~\ref{app:g0}. Upon coarse-graining (i.e. integrating out UV degrees of freedom), the dimensionless coupling constant acquires the logarithmic corrections and becomes scale-dependent $g(\ell)$. This process is described by the standard RG equations, which in one-loop approximation takes the following form \cite{Poboiko_2025}:
\begin{equation}
\label{eq:RG}
\frac{dg}{d\ln\ell}=\beta(g)=-\frac{R}{2\pi}+O(g^{-1}).
\end{equation}
The bare coupling constant $g_0$, given by Eq.~\eqref{eq:g0}, is to be understood as the initial condition for the RG flow equation at the UV scale $g(\ell_0) = g_0$. The situation is then similar to the conventional two-dimensional Anderson localization in orthogonal class: the negative one-loop correction (called ``weak localization'' (WL), in analogy with the theory of Anderson transitions) signals the flow of the system to the strong coupling regime $g(\ell) = O(1)$ (``strong localization''), and eventually to zero in the thermodynamic limit, i.e., $g(\ell \to \infty) \to 0$.
This is manifested in the exponential decay of correlations and an area-law behavior of the entanglement entropy in the thermodynamic limit. The length-scale at which the strong-coupling regime is reached provides an estimate for the ``localization length'':
\begin{equation}
\label{eq:loc-length}
\xi\simeq\ell_{0}\exp\left(2\pi g_{0}\right),
\end{equation}
in the physically relevant replica limit $R \to 1$.

The following comment on the case of very large driving periods, when $\gamma T /2 \gg 1$, is in order here.
In this situation, the two-dimensional space-time splits into strips of large width $T/2$ with alternating values of the hopping: $J_1 = J$ and $J_2 = - J+\epsilon$. For such large $T/2$, one can associate bare effective conductivities $g_{0}^{(1)}= |J_1|/2\gamma$ and  $g_{0}^{(2)}= |J_2| /2\gamma$ with these regions. Correspondingly, the renormalization according to Eq.~\eqref{eq:RG} will first proceed within each of the regions, up to the scale determined by the period $T$. At this scale, the renormalized conductivities will merge into a macroscopic conductivity tensor, thus determining a starting point for a new RG flow that will apply to spatial scales $\ell > T J$.
The separate renormalization of $g_0^{(1)}$ and $g_0^{(2)}$ is particularly important when $|J_1|/\gamma$ and $|J_2|/\gamma$ are very different and one of them is rather small, so that there should be strong localization effects in the corresponding strip already at the scale set by its width $T/2$. We will not consider these regimes in the present paper, limiting ourselves to moderately large $T$ and also comparable $|J_1|$ and $|J_2|$.
Let us emphasize that, also in the case of very large $T$, the field theory at large scales, $\ell > T J$, will have the universal NLSM form of Eq.~\eqref{eq:NLSMAction}, so that all our conclusions on physical observables (see below) apply. It is only the calculation of the coupling constant playing a role of $g_0$ that may require additional efforts in such a case, as explained above.

\subsection{Physical observables}
\label{sec:analytics_observables}

Here we briefly summarize the analytical predictions for the long-distance asymptotic behavior of various observables, which follows from the NLSM field theory as discussed above. Most of these results were obtained in earlier papers \cite{Poboiko_2023,Poboiko_2025}.

The simplest NLSM observable that can be calculated is the density-density correlation function, defined microscopically as follows:
\begin{equation}
{\cal C}_{ij}=\left\langle \hat{n}_{i}\hat{n}_{j}\right\rangle -\left\langle \hat{n}_{i}\right\rangle \left\langle \hat{n}_{j}\right\rangle.
\end{equation}
Its value averaged over quantum trajectories is given by the boundary correlation function of NLSM currents (see Ref.~\cite{Poboiko_2025}). 
After averaging, the correlation function depends only on the relative distance $r = |i-j|$, and its Fourier transform allows one to probe directly the scale-dependent coupling constant:
\begin{equation}
{\cal C}(q)\simeq g(\ell=|q|^{-1})|q|.
\end{equation}
Substituting here the flow behavior of the coupling constant that follows from the one-loop RG form as mentioned in Eq.~\eqref{eq:RG}, one finds \cite{Poboiko_2023}
\begin{equation}
\label{eq:Cq_over_q_weakloc}
{\cal C}(q)\simeq \left(g_{0}-\frac{1}{2\pi}\ln\frac{1}{|q|\ell_{0}}\right) |q|.
\end{equation}
This formula holds in the  ``diffusive'' regime, where $\xi^{-1} \ll q \ll \ell_0^{-1}$.

Such a knowledge of the density correlation function also allows to calculate the variance of charge in a subsystem, which is related to the entanglement entropy via the Klich-Levitov formula \cite{KlichLevitov,Poboiko_2023}:
\begin{equation}
S_{\text{vN}}(\ell)\approx\frac{\pi^{2}}{3}{\cal C}_{2}(\ell),
\end{equation}
with 
\begin{equation}
{\cal C}_{2}(\ell)\equiv\left\langle \hat{N}^{2}_{\ell}\right\rangle -\left\langle \hat{N}_{\ell}\right\rangle ^{2}=\sum^{\ell}_{i,j=1}{\cal C}_{ij}.
\end{equation}
The contribution to the charge variance comes from all length-scales between $\ell_0$ and $\ell$; and as a consequence, it is also related to the running coupling constant via the following integral \cite{Poboiko_2023}:
\begin{equation}
{\cal C}_{2}(\ell)\approx\frac{2}{\pi}\int^{\ln\ell}_{\ln\ell_{0}}g(\ell)d\left(\ln\ell\right).
\label{eq:C2-integral}
\end{equation}
Here, the numerical prefactor includes the factor of two, which comes from two cuts which separate the subsystem from its complement.
This gives the following prediction for the behavior of the entanglement entropy in the ``diffusive'' regime $\ell \ll \xi$:
\begin{equation}
S_{\text{vN}}(\ell\ll\xi)\simeq\frac{2\pi}{3}\left(g_{0}\ln\frac{\ell}{\ell_{0}}-\frac{1}{4\pi}\left(\ln\frac{\ell}{\ell_{0}}\right)^{2}\right).
\label{Eq:S_vn(l)}
\end{equation}
Furthermore, by taking $\ln \xi$ as an upper limit of the integral in Eq.~\eqref{eq:C2-integral}, we can obtain the area-law saturation value of the charge variance and of the entropy at $\ell \gg \xi$:
\begin{equation}
\label{eq:entropy-saturated}
S_{\text{vN}}(\ell\to\infty)\equiv S^{\text{sat}}_{\text{vN}}\simeq\frac{2\pi^{2}}{3}g^{2}_{0}\,.
\end{equation}
We will compare the analytical findings with numerical results in Sec.~\ref{Sec:Numerical_results}.

\subsection{Ballistic-diffusive crossover}
\label{Subsec:WH_integral}

The NLSM describes the large-distance behavior, $\ell \gg \ell_0$, which corresponds, in momentum space, to $q \ll \ell_0^{-1}$. This includes both the diffusive and the localized regimes.
The opposite limit of short scales, $\ell \ll \ell_0$ or $q \gg \ell_0^{-1}$ corresponds to the ballistic regime, where the effects of monitoring play only a minor role. In this subsection, we perform an analysis of the crossover between the ballistic and diffusive regimes in an approximation that neglects WL loop corrections. Within this approximation, the analytical description of the crossover is based on summation of Feynman diagrams of ladder type.
The detailed derivation is presented in Appendix~\ref{app:crossover}.  Here, we briefly summarize the key points of the calculation. Let us emphasize that, at variance with the diffusive regime that is universally described by the NLSM as given in Eq.~\eqref{eq:NLSMAction} with a single coupling constant, the ballistic-diffusive crossover depends on the microscopic model. 

The first step is the evaluation of the average fermionic Green's function, for which the self-consistent Born approximation is exact \cite{Poboiko_2023}. Since the Hamiltonian given in Eq.~\eqref{Eq:Hamiltonian} is translationally invariant in space, the Green's function can be calculated exactly in a mixed momentum-time representation, yielding:
\begin{equation}
\label{eq:GR}
iG^{R}_{k}(t,t^{\prime})=\Theta(t-t^{\prime})\exp\left(-i\int^{t}_{t^{\prime}}\varepsilon_{k}(\tau)d\tau\right)\exp\left(-\frac{t-t^{\prime}}{2\tau_{0}}\right).
\end{equation}
Here $\tau_0 = \gamma^{-1}$ is the mean-free time and $\varepsilon_k(\tau) = 2 J(\tau)\cos k$ is the instantaneous quasiparticle dispersion relation.

The main building block of the diffuson ladder diagrams is the loop involving retarded and advanced Green's functions. In the same mixed momentum-time representation, it can also be calculated exactly, with the following result:
\begin{align}
\label{eq:WH:Bq}
{\cal B}_{q}(t,t^{\prime})&=\tau_0^{-1} \int^{\pi}_{-\pi}\frac{dk}{2\pi}G^{R}_{k+q/2}\left(t,t^{\prime}\right)G^{A}_{k-q/2}\left(t^{\prime},t\right)\notag\\
&=\tau_0^{-1} \Theta(t-t^{\prime})\exp\left(-\frac{t-t^{\prime}}{\tau_{0}}\right)J_{0}\left(2\widetilde{q}\int^{t}_{t^{\prime}}J(\tau)d\tau\right).
\end{align}
Here, $\widetilde{q} = 2 \sin(q/2)$ is the rescaled momentum which incorporates also the finite size of the Brillouin zone, and $J_0(z)$ is the Bessel function of order zero.

The diffuson ladder is then given by the geometric series, and its calculation requires inversion of the corresponding integral operator:
\begin{equation}
\label{eq:integral-operator-Dq}
\hat{{\cal D}}_{q}=\left(1-\hat{{\cal K}}_{q}\right)^{-1},\quad\hat{{\cal K}}_{q}=\hat{{\cal B}}_{q}\hat{{\cal B}}^{T}_{q},
\end{equation}
where $\hat{{\cal B}}_{q}$ is an integral operator acting in the time domain with the kernel ${\cal B}_{q}(t,t^\prime)$. Importantly,  all the involved operators are restricted to the time domain $t<0$, where $t=0$ is the time at which steady-state observables are calculated, where the 
 evolution is started at time $t_i \to -\infty$. This restriction to the $t<0$ domain breaks the translational invariance in time, thus greatly complicating the analytic inversion of the operator in Eq.~\eqref{eq:integral-operator-Dq}.

The diffuson ladder $\hat{{\cal D}}_{q}$ determines the behavior of the density-density correlation function operator given by:
\begin{equation}
\label{eq:Cq-operator}
\hat{{\cal C}}_{q}=\frac{1}{4\gamma}\left[1-\left(1-\hat{{\cal B}}^{T}_{q}\right)\hat{{\cal D}}_q\left(1-\hat{{\cal B}}_{q}\right)\right].
\end{equation}
The physical density-density correlation function is then given by the boundary value of this operator as:
\begin{equation}
\label{eq:CqWH}
{\cal C}_0(q)=\lim_{t,t^{\prime}\to-0}{\cal C}_{q}(t,t^{\prime}).
\end{equation}
Here the subscript ``0'' indicates that the correlation function  ${\cal C}(q)$ is evaluated in the ladder approximation (or, equivalently, Gaussian approximation for the NLSM functional integral), which discards loop corrections. 

As discussed above, the major difficulty for evaluation of $\hat{{\cal C}}_{q}$ as given by  Eq.~\eqref{eq:Cq-operator} is related to the necessity to invert an integral operator with time translation invariance broken by the restriction $t < 0$. This also makes it difficult to estimate $\hat{{\cal C}}_{q}$ numerically.
However, only the boundary value of operator $\hat{{\cal C}}_q$ (see Eq.~\eqref{eq:CqWH}) is physically relevant. As a consequence, the problem can be greatly simplified: instead of inverting the operator $(1 - \hat{{\cal K}}_q)$, one can solve the following Wiener-Hopf-type integral equation for a function $F_q(t)$:
\begin{equation}
\label{eq:WH:Fq}
F_{q}(t)-\int^{0}_{-\infty}dt^{\prime}{\cal K}_{q}(t,t^{\prime})F_{q}(t^{\prime})={\cal K}_{q}(t,0),
\end{equation}
with 
\begin{equation}
\label{eq:WH:Kq}
{\cal K}_{q}(t,t^{\prime})=\int^{0}_{-\infty}dt^{\prime\prime}{\cal B}_{q}(t,t^{\prime\prime}){\cal B}_{q}(t^{\prime},t^{\prime\prime}).
\end{equation}
The physical density-density correlation function is then immediately obtained from the solution of this equation as
\begin{equation}
\label{eq:WH:Cq}
{\cal C}_0(q)=\frac{1}{4}\left(1- \gamma^{-1}F_{q}(0)\right).
\end{equation}
At large $q$, the $F_q(t)\to 0$ and ${\cal C}_0(q) \to 1/4$. Equations (\ref{eq:WH:Bq}), (\ref{eq:WH:Fq}), \eqref{eq:WH:Kq}, and \eqref{eq:WH:Cq} provide a closed set of equations which determine the density-density correlation functions ${\cal C}_0(q)$
in the ballistic-diffusive crossover around $q \sim \ell_0^{-1}$, in the approximation neglecting the WL correction.  

In the diffusive limit, $q\ell_0 \ll 1$, the correlation function ${\cal C}_0(q)$ takes the diffusive form, ${\cal C}_0(q) \simeq g_0 |q|$ [analogous to Eq.~\eqref{eq:Cq_over_q_weakloc} but without the WL correction].
At the same time, for values of $q\ell_0$ that are in the diffusive regime but not extremely small (such as $10^{-2} \lesssim q\ell_0 < 1$), which are relevant to numerical simulations, a correction to the diffusive asymptotics of ${\cal C}_0(q)$ originating from the ballistic-diffusive crossover may still be noticeable and comparable to the WL correction. In view of this, we will use the full form of ${\cal C}_0(q)$ (and not simply its diffusive asymptotics) when determining the WL correction $\delta{\cal C}(q) = {\cal C}(q) - {\cal C}_0(q)$ from the numerical data of the density correlation function in Sec.~\ref{sec:numerics_Cq_weakloc}.

\section{Numerical results}
\label{Sec:Numerical_results}

In this section, we support and complement our analytical findings discussed in Sec.~\ref{Sec:Analytic} by direct numerical simulations of quantum trajectories in the model under consideration. 
For the unitary part of the evolution, which is 
described by the Hamiltonian given in Eqs.~\eqref{Eq:Hamiltonian} and \eqref{Eq:Jt}, we consider the following two cases: (i) $\epsilon=0$, which corresponds to the ``maximally symmetric'' drive, with $J(t)$ having the same magnitude (but opposite signs) in two halves of the period, and (ii) $\epsilon=3J/4$ (which is the value on which Ref.~\cite{Modak_2025} focused to present arguments for the BKT transition). The monitoring protocol is described in Sec.~\ref{Sec:Model} and in Appendix~\ref{app:SSE}.

Our numerical studies focus on
the (steady-state) von-Neumann entanglement entropy and the density-density correlation in momentum space, evaluated at the stroboscopic periods $t=nT$.

\begin{figure}
\centering
    \includegraphics[width=0.49\textwidth,height=0.4\textwidth]{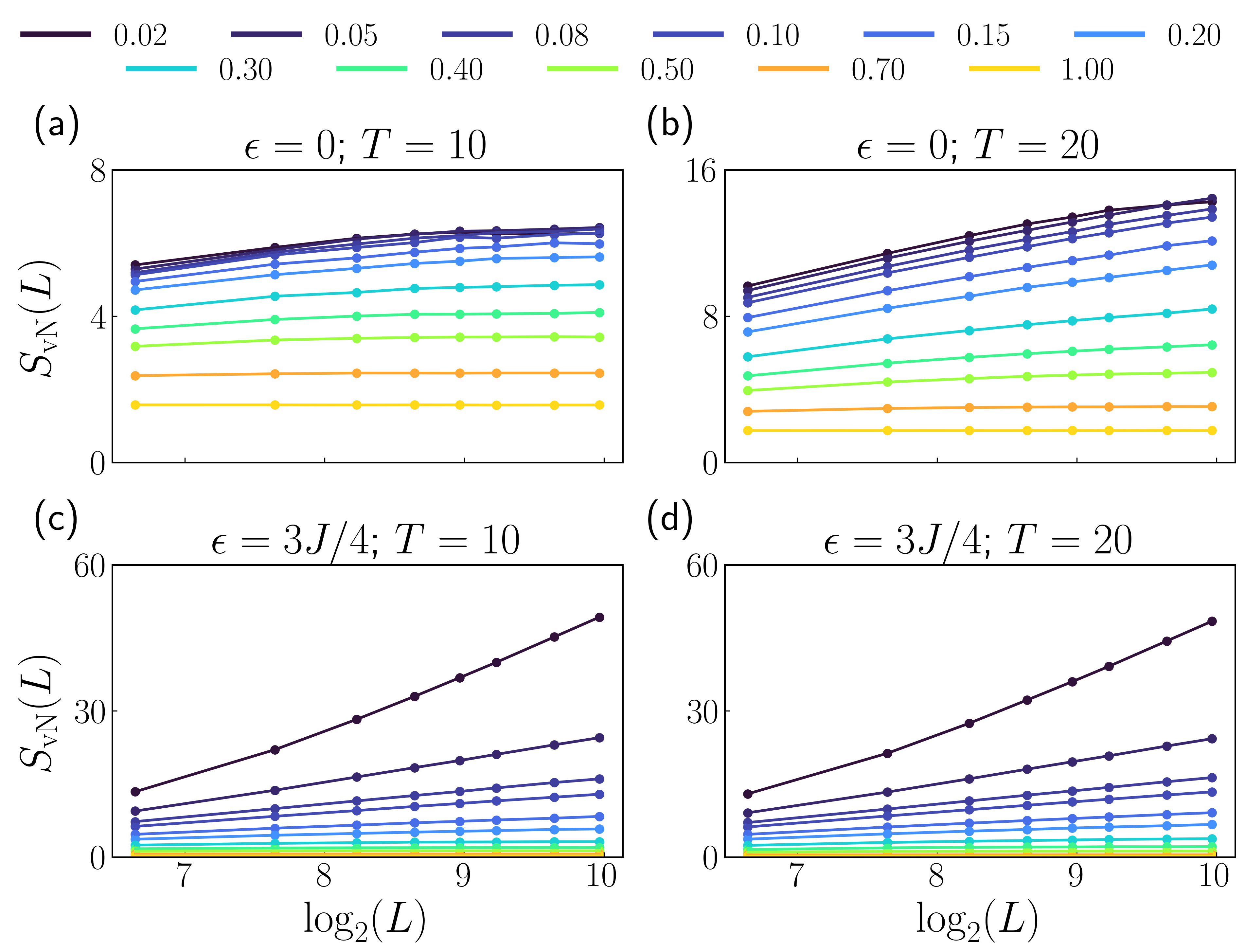}
    \caption{Average half-chain entanglement entropy $S_{\mathrm{vN}}$ as a function of system size $L$.
  Panels (a,b) correspond to the maximally symmetric driving ($\epsilon=0$) with period $T=10$ and $T=20$, respectively, while panels (c,d) correspond to the case $\epsilon=3J/4$ with the same driving periods. The system sizes considered here are $L=100,200,300,400,500,600,800,$ and $1000$. For $L\leq600$, all possible half-chain cuts for every state are included in the averaging, whereas for $L=800$ and $L=1000$, the averages are taken over $20$ approximately equidistant cuts. In addition, the averaging over quantum trajectories and over the time in the steady-state regime is performed, see text in Sec.~\ref{Subsec:Entanglement_entropy} and in Appendix~\ref{app:SSE}. Different colors correspond to different values of the measurement rate $\gamma/J$, see legend.}
    \label{Fig:Fig_2}
\end{figure}

\subsection{Entanglement entropy}
\label{Subsec:Entanglement_entropy}

The von Neumann entropy of a subsystem is a commonly used measure to characterize entanglement. It serves as a sensitive probe to characterize the phases (or regimes) in the context of MIPTs. The entropy of a subsystem A is defined as $S_{\text{vN}}^{(A)}=-\text{Tr}(\rho_A \text{ln}\rho_A$) \cite{Amico_2008}, where $\rho_A$ is the reduced density matrix of A obtained by tracing out the degrees of freedom of the rest of the subsystem B representing the rest of the system,  i.e., $\rho_A=\text{Tr}_B \ket{\Psi_t}\bra{\Psi_t}$. For a Gaussian system (as studied in this paper), the von Neumann entropy can be expressed in terms of eigenvalues $\lambda_i^{(A)}$ of the equal-time correlation function $\langle c_{i}^{\dagger} c_{j}\rangle$ in the subsystem A \cite{Calabrese_2005,Peschel_2009}:
\begin{equation}
S_{\mathrm{vN}}^{(A)}=-\sum_{i=1}^{l}\left[\lambda_i^{(A)}\ln\lambda_i^{(A)}+\left(1-\lambda_i^{(A)}\right)\ln\left(1-\lambda_i^{(A)}\right)\right].
\label{Eq:S_vN}
\end{equation}
Upon averaging over the ensemble of quantum trajectories, $S_{\mathrm{vN}}$ becomes a function of the size $L$ of the whole system and size $l_A$ of the subsystem. Since we will analyze below only the (most frequently considered) case of half-chain entropy ($l_A = L/2$), we will keep only $L$ as an argument. 

To determine numerically the steady-state entanglement entropy, we first inspect its time dependence and identify the saturation regime. For all parameters ($\epsilon$ and $T$) and at all values of $\gamma$ considered in this work, the entropy reaches the plateau no later than at time $t=1200$.  Therefore, we perform the time averaging over the interval $1200 \leq t \leq 2000$, corresponding to the steady-state plateau regime. This time averaging is performed in addition to the averaging over 50 quantum trajectories and over different positions of the cuts dividing the system into two subsystems of length $L/2$. 

The data for the length dependence of the average entropy are shown in Fig.~\ref{Fig:Fig_2}, in the range $100 \le L \le 1000$ (note that the scale of the $L$ axis is logarithmic, so that $\ln L$ behavior is represented by a straight line).
An inspection of the data indicates a consistency with our analytical predictions based on NLSM formalism, see Sec.~\ref{Sec:Analytic}. 
For large values of $\gamma$ (for either $\epsilon=0$ or $3J/4$), the entropy is fully saturated already at $L=100$, so that the system is clearly in the area-law phase with a small localization length $\xi$
 as expected. For smaller values of $\gamma$, the system  first exhibits a logarithmic growth of the entropy at intermediate $L$, followed by its saturation (or at least a clear tendency to saturation) at larger $L$. For $\epsilon=0$, this tendency to saturation is obvious even for smallest $\gamma$. This behavior is also in agreement with analytical results of Sec.~\ref{Sec:Analytic}, which predict a logarithmic growth with a WL correction in the diffusive regime, $l_0 \ll L \ll \xi$, followed by strong localization at $L > \xi$. At the same time, for $\epsilon=3J/4$ and the smallest values of the measurement rate ($\gamma= 0.02,\:0.05$), the tendency to saturation is not seen in the data and the curves even exhibit an upward bending (especially pronounced for $\gamma=0.02$). This behavior is consistent with analytical expectations. Indeed, these data sets correspond to rather large values of the bare conductivity $g_0$, see Fig.~\ref{Fig:Fig_1}, which implies large values of the mean free path $l_0$ and exponentially large values of the localization length $\xi$. Correspondingly, at system sizes accessible in our numerics, the results are affected by a crossover from the  ballistic (volume law) to the diffusive (logarithmic law) regime, explaining the observed upward bending. 

 Clearly, if one relies solely on numerical data like those shown in the panels (c) and (d) of Fig.~\ref{Fig:Fig_2}, one may be tempted to conclude that there is a thermodynamic-limit ($L\to\infty$) phase transition between the area-law phase at larger $\gamma$ and a logarithmic phase at smaller $\gamma$.  Indeed, such a conjecture was made
in Ref.~\cite{Modak_2025} who interpreted their numerical results in terms of a BKT transition. We are armed, however, with analytical predictions (Sec.~\ref{Sec:Analytic}), according to which there is no transition in the sense that the system is in the area-law phase at $L\to\infty$ for any finite $\gamma$. Of course,  reaching numerically the asymptotic area-law  behavior becomes impossible for sufficiently large $g_0$, in view of exponential divergence of the localization length $\xi$. Nevertheless, we will  present a strong numerical support of the analytical results by performing a quantitative comparison between the analytical predictions and numerical findings. Specifically, we demonstrate below a very good agreement between the analytical and numerical results with respect to the bare conductivity, the WL correction, and the saturation value of the entanglement entropy, as presented in Secs.~\ref{Subsubsec:g_0_comparison}, \ref{Subsubsec:WL}, and \ref{Subsubsec:SL}, respectively.

\begin{figure}
\centering
\includegraphics[width=0.49\textwidth,height=0.25\textwidth]{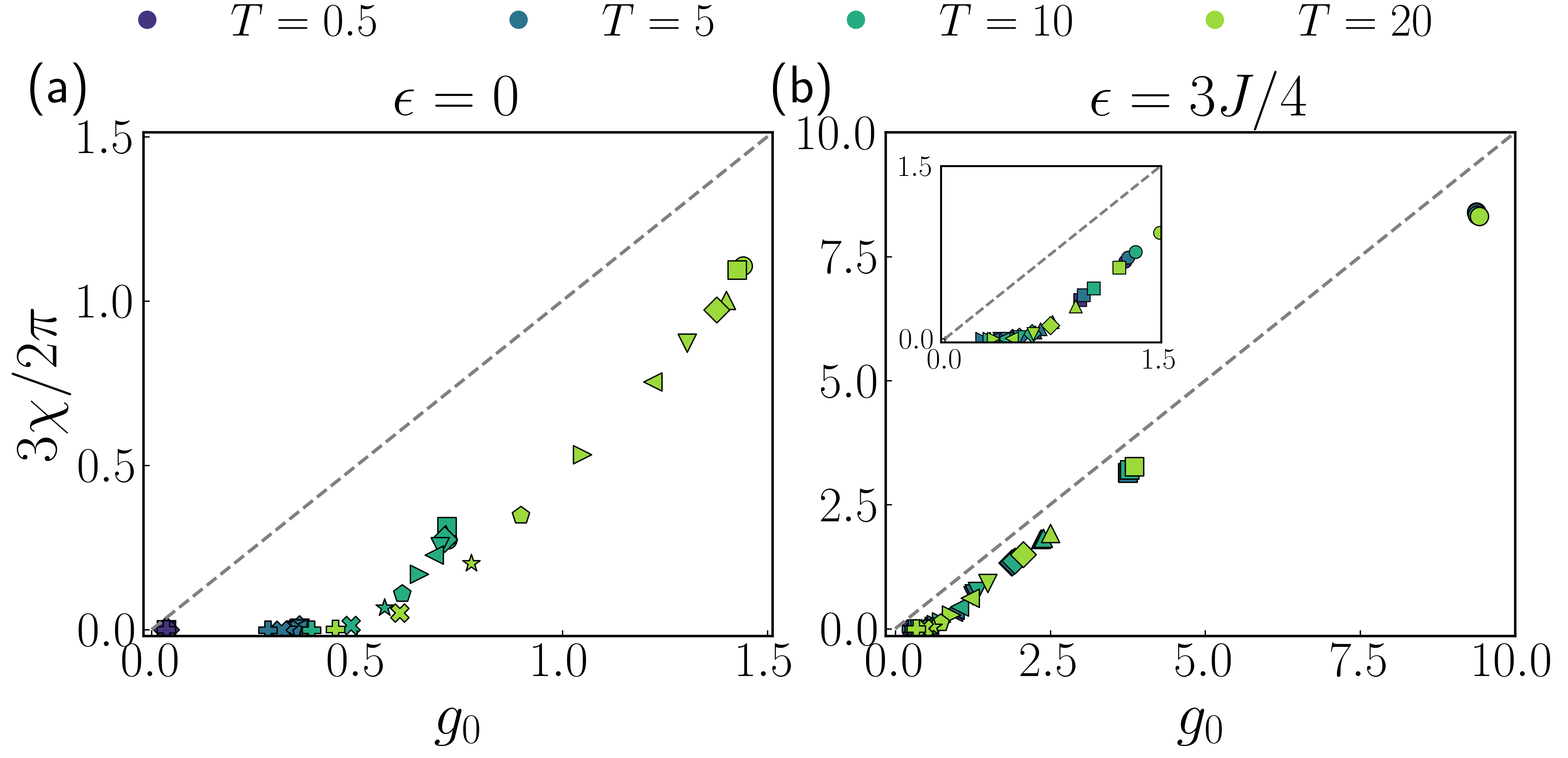}
    \caption{Numerically determined coefficient $\chi$ (multiplied by $3/2\pi$) of the initial logarithmic growth of the entanglement entropy vs the analytical value of the bare diffusion constant $g_0$, Eq.~\eqref{eq:g0}, for (a) $\epsilon=0$ and (b) $\epsilon=3J/4$. The data for different time periods (see legend) and for different values of the measurement rate $\gamma/J$
    (identical to those in Fig.~\ref{Fig:Fig_2}) are included. The inset of panel (b) provides a zoom on the region of small $g_0$.
    }
    \label{Fig:Fig_3}
\end{figure}

\subsubsection{Bare conductivity}
\label{Subsubsec:g_0_comparison}

According to the analytical predictions in Eq.~(\ref{Eq:S_vn(l)}), the derivative $\chi$ of the entanglement entropy with respect to $\ln L$ is proportional to the ``conductivity'' $g$ at scale $L$:
\begin{equation}
\label{eq:dS_d_ln_L}
\chi(L) \equiv \frac{\partial S_{\rm vN}(L)}{\partial \ln L} = \frac{2\pi}{3} g(L) \,,
\end{equation}
with
\begin{equation}
\label{eq:gL-weakloc}
g(L) \simeq g_0 - \frac{1}{2\pi} \ln (L/\ell_0) \,.
\end{equation}
These formulas are controllable in the diffusive regime, which implies $g_0 \gg 1$ and $\ell_0 \ll L \ll \xi$.  For $L$ in the initial part of the diffusive regime, and for a sufficiently large  $g_0$, the WL correction in
Eq.~\eqref{eq:gL-weakloc} is small compared to $g_0$, so that the derivative $\chi$ in Eq.~\eqref{eq:dS_d_ln_L} should give the bare conductivity $g_0$ times $2\pi/3$. 
We are now going to confront our numerical results with this analytical prediction. 

For $L$ to be unambiguously in the diffusive regime (rather than in the ballistic-to-diffusive crossover), it should exceed sufficiently well the scale $\ell_0$. In practice, the condition $L \gtrsim 10 \ell_0$ is reasonably sufficient. For nearly all  parameters that we used, the range of our smaller length scales, $L = 100 - 300$ satisfies this condition and can be used for numerical estimate of the initial value of $\chi$. 

In Fig.~\ref{Fig:Fig_3}, we plot the slope $\chi$  characterizing the initial part of the diffusive regime in curves shown in Fig.~\ref{Fig:Fig_2} (for various $\epsilon$, $T$, and $\gamma$)
versus the analytical value of the bare conductivity $g_0$ obtained in Eq.~\eqref{eq:g0}. We also include the data for $T=0.5$ and $T=5$. 
The data for $\chi$ are multiplied by $3/2\pi$ to simplify comparison with the prediction in Eq.~\eqref{eq:dS_d_ln_L}. As illustrated in Fig.~\ref{Fig:Fig_3}, we observe a very good agreement, up to a downward offset by $\sim 0.5$. This agreement is rather remarkable, taking into account that the numerical values of $g_0$ are not so large. The offset is expected since there is necessarily a quantum correction of order unity to the conductivity $g_0$ originating from the ballistic-diffusive crossover (or, equivalently, from the early part of the diffusive regime).  As a rough estimate, one can substitute $L=10 \ell_0$ into Eq.~\eqref{eq:gL-weakloc}, which will yield a shift $g(L)-g_0 \approx - 0.4$, comparable to the one that is observed in 
Fig.~\ref{Fig:Fig_3}.

For the smallest values of $g_0$, i.e., for $g_0 \lesssim 0.5$, the numerical data show the zero slope $\chi$. We have a clear explanation for this as well. For such small values of $g_0$, the localization length $\xi$ is short, so that systems with $L\ge 100$ are in the strongly localized regime, implying the area-law behavior of the entanglement entropy, i.e., its $L$-independence.
The apparent kink that is observed at $g_0\approx 0.5$ is thus a manifestation of a crossover from strong to WL as observed at a scale $L \simeq 100$. If the system is studied at larger length scales, the position of this crossover will move towards larger values of $g_0$, as predicted by Eq.~\eqref{eq:gL-weakloc}.
We are now going to demonstrate that the $L$-dependence of the slope of the entropy $\chi(L)$ (as given in Eq.~\eqref{eq:dS_d_ln_L}), indeed follows the analytical prediction for the conductivity renormalization induced by WL as obtained in Eq.~\eqref{eq:gL-weakloc}. 

\begin{figure}
\centering
   \includegraphics[width=0.492\textwidth,height=0.35\textwidth]{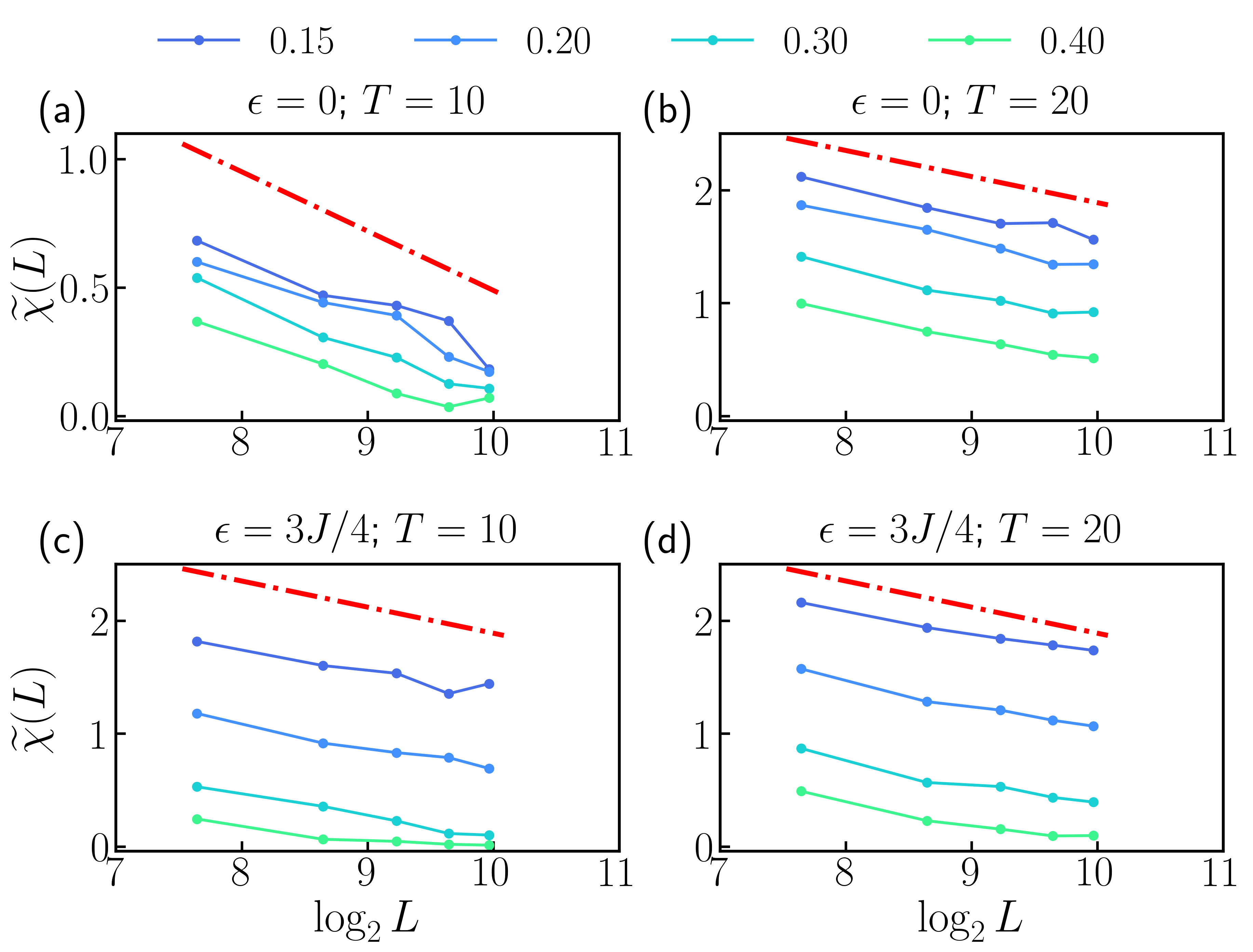}
    \caption{Discrete version of the entanglement entropy derivative in Eq.~\eqref{eq:dS_d_ln_L}, defined according to Eq.~\eqref{Eq:Discrete_derivative} by using the data presented in Fig.~\ref{Fig:Fig_2}. Different colors correspond to different values of the measurement rate $\gamma/J$ (see legend), which are chosen in the range from 0.15 to 0.4 providing access to the diffusive regime for the numerically studied system sizes $L$. The red dash-dotted straight lines have the slope $-1/3$  visualizing the analytically predicted WL renormalization as given in Eq.~\eqref{eq:chi_L_weakloc}. }
    \label{Fig:Fig_4}
\end{figure}

\subsubsection{Weak localization}
\label{Subsubsec:WL}

The logarithmic WL correction to the conductivity, Eq.~\eqref{eq:gL-weakloc}, is a hallmark of the NLSM universality. The coefficient $1/2\pi$ in front of the logarithm is a universal number determined solely by the BDI symmetry class of the problem. Thus, a numerical verification of this logarithmic renormalization with the predicted coefficient provides additional support to the analytical theory. 

According to Eqs.~\eqref{eq:dS_d_ln_L} and \eqref{eq:gL-weakloc}, the entropy derivative $\chi(L)$, should show, in the diffusive regime, the following logarithmic dependence on $L$, 
\begin{equation}
\label{eq:chi_L_weakloc}
\chi(L) = \chi_0 - \frac{1}{3} \ln(L/\ell_0) \,.
\end{equation}
To verify this WL behavior, we plot in Fig.~\ref{Fig:Fig_4} a discrete version of the derivative $\chi(L)$ (Eq.~\eqref{eq:dS_d_ln_L}), defined as \cite{Fava_2023,FavaNahum2024}

\begin{eqnarray}
    \widetilde{\chi}(L)= \frac{S_{\text{vN}}(L)- S_{\text{vN}}(L/2)}{\text{ln}~2}.
 \label{Eq:Discrete_derivative}
\end{eqnarray}
In this plot, the values of the measurement rate $\gamma/J$ are in the intermediate range (between 0.15 and 0.4), so that the system is in the diffusive regime. The logarithmic renormalization of $\chi(L)$ is clearly seen in the plot. Moreover, the slope of this renormalization is in a very good agreement with the analytically predicted coefficient $-1/3$. The fact that this behavior is observed in the data for different values of $\epsilon$, $T$, and $\gamma$ nicely confirms the universality of the WL renormalization predicted by NLSM. Furthermore, the agreement of the slope in Fig.~\ref{Fig:Fig_4} with the predicted value $-1/3$ demonstrates that the monitored free-fermion system remains in the BDI universality class in the presence of a time-periodic drive since the drive preserves the underlying symmetries of the problem.

\begin{figure}
\centering
\includegraphics[width=0.447\textwidth,height=0.265\textwidth]{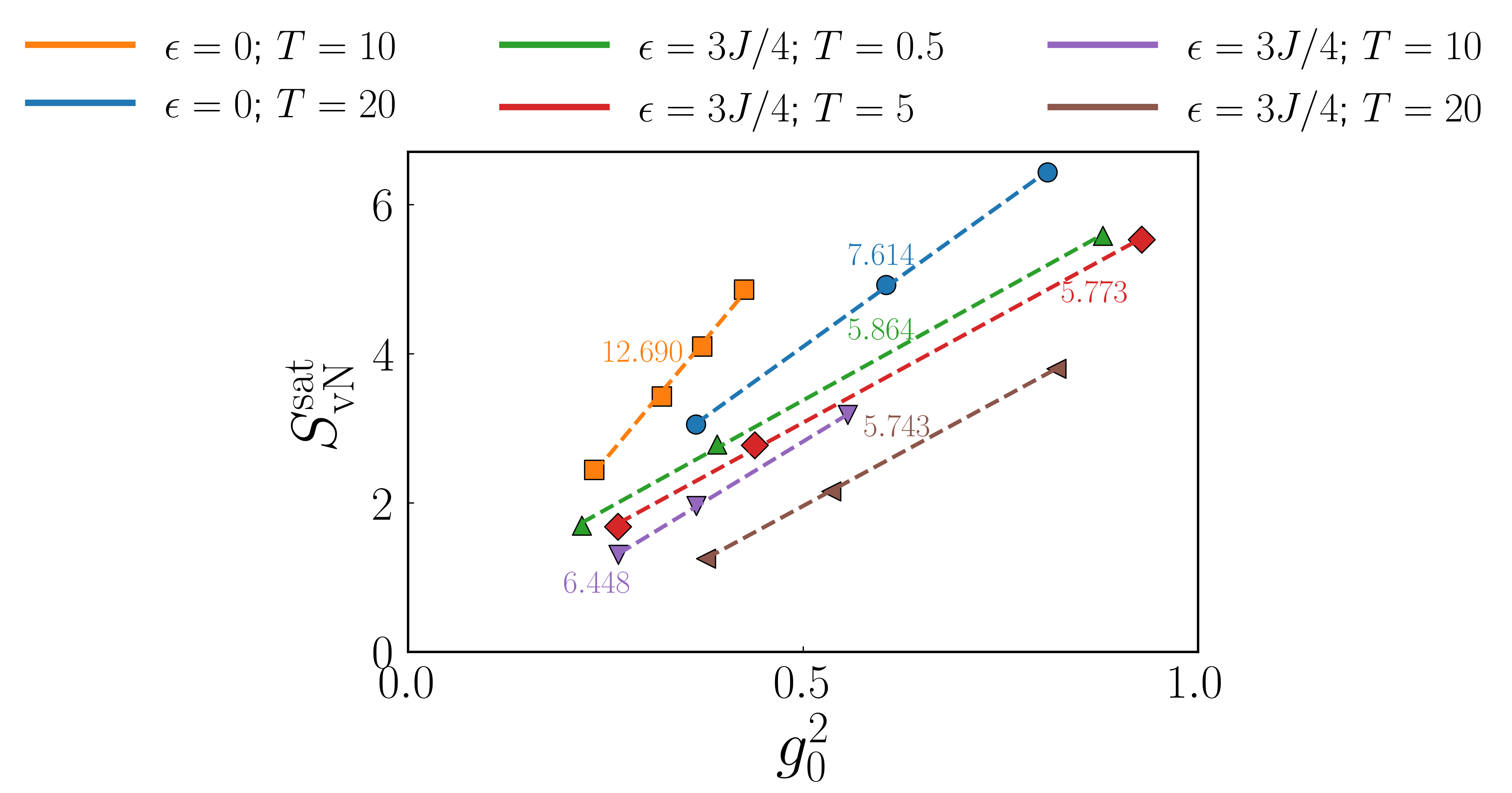}
    \caption{Numerical verification of the linear relation in Eq.~\eqref{eq:entropy-saturated-2} between the saturated value $S_{\rm vN}^{\rm sat}$ of the entanglement entropy at $L\gg \xi$ and the square of the bare conductivity $g_0$. The values of $S_{\rm vN}^{\rm sat}$ are obtained from those data of Fig.~\ref{Fig:Fig_2} that exhibit saturation at numerically studied system sizes. The values of $g_0$ are given by Eq.~\eqref{eq:g0}. Different colors correspond to different parameters $(\epsilon, T)$ of the time-periodic Hamiltonian ${\cal H}(t)$, Eq.~\eqref{Eq:Hamiltonian}. Different points for each color correspond to different values of the measurement rate $\gamma/J$. The slopes corresponding to each data set with given  ${\cal H}(t)$ and varying $\gamma/J$ are indicated and visualized by dashed lines. The analytically predicted value of the slope according to Eq.~\eqref{eq:entropy-saturated-2}    is $2\pi^2/3 \approx 6.6$. 
 }
    \label{Fig:Fig_5}
\end{figure}

\subsubsection{Strong localization: Saturation value}
\label{Subsubsec:SL}

In the strong-localization regime, $L \gg \xi$, the entanglement entropy saturates (becomes $L$-independent), which is the typical area-law behavior. The saturation value is predicted to be [see Eq.~\eqref{eq:entropy-saturated}]
\begin{equation}
S_{\rm vN}^{\rm sat} \simeq \frac{2\pi^2}{3}g_0^2 \,.
\label{eq:entropy-saturated-2}
\end{equation}
It is worth reiterating that, in analogy to other NLSM predictions,  the analytical result in Eq.~\eqref{eq:entropy-saturated-2}
is controllable for $g_0 \gg 1$.

While such a saturation is predicted for arbitrary parameters of the Hamiltonian and measurement rate, its numerical verification is restricted to systems with rather modest values of $g_0$, in view of the exponential increase of the localization length, Eq.~\eqref{eq:loc-length}. In Fig.~\ref{Fig:Fig_5}, we show the values of $S_{\rm vN}^{\rm sat}$ obtained from those data of Fig.~\ref{Fig:Fig_2} (and analogous data for $T=0.5$ and $T=5$) that exhibit saturation in the entropy. 
To verify the relation obtained in Eq.~\eqref{eq:entropy-saturated-2}, the saturation values are plotted versus $g_0^2$, with $g_0$ given by Eq.~\eqref{eq:g0}, where different colors correspond to different parameters $(\epsilon, T)$ of the time-periodic Hamiltonian ${\cal H}(t)$ in Eq.~\eqref{Eq:Hamiltonian}. 
For each Hamiltonian parameter set $(\epsilon, T)$, there are several values of the measurement rate $\gamma/J$, for which the saturation of the entropy is reached. For each such set, we observe a clear linear dependence of $S_{\rm vN}^{\rm sat}$ on $g_0^2$, as predicted. Furthermore, for almost all parameter sets, the slopes of this linear dependence 
are in the range between 5.7 and 7.6, rather close to the analytically predicted value $2\pi^2/3 \approx 6.6$. (The only exception with a considerably larger slope is provided by the set of parameters $(\epsilon=0, \:T=10)$ that corresponds to the strongest localization, i.e., particularly small values of $g_0^2$.)

The observed agreement is rather remarkable, taking into account that the analytical result is controllable at large $g_0$, whereas the numerics for $S_{\rm vN}^{\rm sat}$ is restricted to systems with $g_0 \le 1$. 

This completes our analysis of numerical results for the entanglement entropy $S_{\rm vN}(L)$, which indeed demonstrates a very good agreement with analytical predictions. Below, we complement (and further support) these results by an analysis of the density correlation function, which is a primary quantity characterizing a Gaussian (free-fermion) system.

\begin{figure*}
\centering
\includegraphics[width=0.99\textwidth,height=0.62\textwidth]{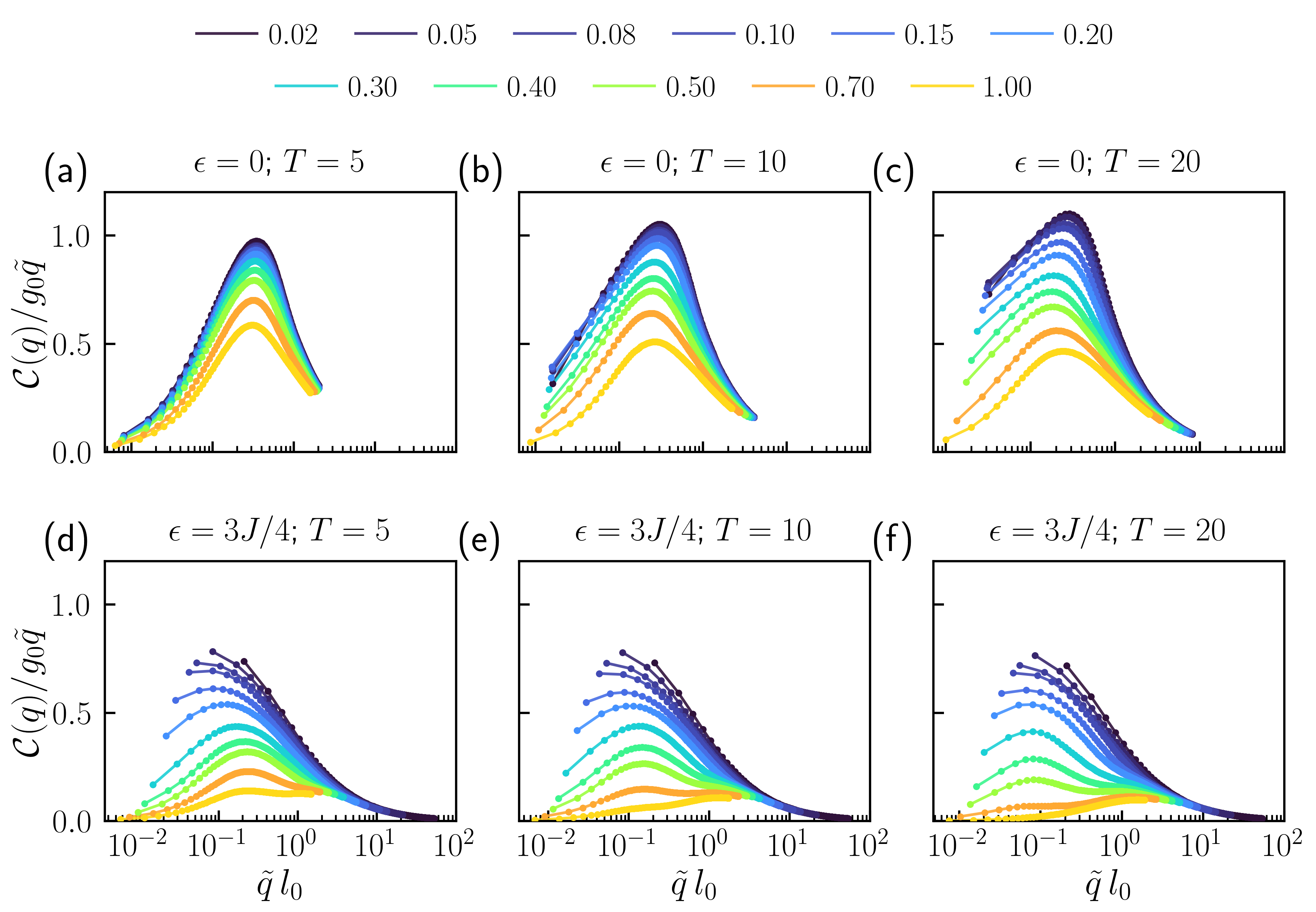}
    \caption{Rescaled density correlation function in momentum space, $\mathcal{C}(q)/(g_0\tilde{q})$, plotted as a function of $\tilde{q}l_0$, where $\tilde{q}= 2 \sin(q/2)$.
  Different colors correspond   
   to different  measurement strengths $\gamma/J$, see legend.   Panels (a–c) correspond to the perfectly symmetric driving ($\epsilon=0$) with periods $T=5,10,$ and $20$, respectively, whereas panels (d–f) correspond to  $\epsilon=3J/4$ with the same driving periods. The system size is taken to be $L=800$, and the correlation function is evaluated at the time $t=1500$.}
    \label{Fig:Fig_6}
\end{figure*}

\subsection{Density correlation function}
\label{Subsec:Density_correlation}

Following the approach of Ref.~\cite{Poboiko_2023}, we study numerically the pair density correlation function (see Secs.~\ref{sec:analytics_observables} and \ref{Subsec:WH_integral}), 
\begin{equation}
    \mathcal{C}_{i,j}=\left\langle \hat n(i) \hat n(j)\right\rangle-\left\langle \hat n(i) \right\rangle \left\langle\hat n(j)\right\rangle = \mathcal{G}_{i,j} \delta_{i,j} - \mathcal{G}_{i,j}\mathcal{G}_{j,i}.
   \label{Eq:Density_correlation}
\end{equation}
Within the analogy with Anderson localization in disordered systems, this quantity is a counterpart of the two-point conductance between the sites $i$ and $j$ at the sample boundary. Performing an averaging over quantum trajectories and Fourier transformation to the momentum space, one obtains the correlation function ${\cal C}(q)$. For numerical analysis, it is convenient to use the ratio ${\cal C}(q) / |q|$. Let us remind its key properties as predicted analytically  (see Sec.~\ref{sec:analytics_observables} for detail):

\begin{itemize}

\item
In the ballistic regime, i.e., at large momenta ($|q|\ell_0\gg 1$), the ratio ${\cal C}(q) / |q|$ scales as $1/|q|$.

\item
In the diffusive regime, ${\cal C}(q) / |q|$ is given by the conductivity $g(L)$ at the scale $L\sim q^{-1}$. In other words, it exhibits the logarithmic WL behavior, Eq.~\eqref{eq:Cq_over_q_weakloc}, with the universal coefficient $-1/2\pi$ for the BDI symmetry class. 

\item
The strong-localization behavior in the large-$L$ limit implies the limiting behavior $\lim_{q\to 0} {\cal C}(q) / |q| = 0$. To observe numerically this behavior, one should access momenta $q < \xi^{-1}$, which requires reaching system sizes $L > \xi$. 

\item
If the system remains diffusive in the thermodynamic limit, the limiting value $\lim_{q\to 0} {\cal C}(q) / |q|$ is equal to the corresponding conductivity $g(L\to \infty)$. While the NLSM prediction is that this does not happen in a $d=1$ system, such behavior is observed on the ``metallic'' side of the MIPT in $d>1$ dimensions \cite{Poboiko_2024}.  

\end{itemize}

We will now compare these predictions with the numerical data.  

\subsubsection{Overall behavior of ${\cal C}(q)$:  Ballistic, WL, and strong-localization regimes}
\label{sec:numerics_Cq_overall}

In Fig.~\ref{Fig:Fig_6}, we show the results for the ratio $\mathcal{C}(q)/g_0 \tilde q$, where 
$\tilde q = 2~ \text{sin}(q/2)$, takes into account \cite{Poboiko_2023} that we deal with a lattice model, with Brillouin-zone periodicity in momentum space. For small momenta, which is the region of our main interest here, the difference between $\tilde q$ and $q$ is immaterial.
If there were no localization corrections, the ratio $\mathcal{C}(q)/g_0 q$ would tend to unity in the limit $q\to 0$, see
Eq.~\eqref{eq:Cq_over_q_weakloc}. Thus, in the absence of localization, the $1/q$ behavior of  $\mathcal{C}(q)/g_0 q$ at large momenta would saturate with lowering $q$, crossing over to 
$\mathcal{C}(q)/g_0 q \to 1$ at $q\to 0$. The localization should show up as a non-monotonic behavior, with $\mathcal{C}(q)/g_0 q$ reducing towards zero when $q$ is lowered. 

For convenience, we plot the results in Fig.~\ref{Fig:Fig_6} (and also in Fig.~\ref{Fig:Fig_7} below) as  functions of $\tilde{q}\ell_0$, with $\ell_0 = 2\sqrt{2} g_0$.

The six panels of Fig.~\ref{Fig:Fig_6} correspond to different parameters $(\epsilon, T)$ of the driving; in each of them, the measurement rate $\gamma/J$ is varied from 0.02 to 1.00.  For larger values of the measurement rate, the data clearly exhibit the strong-localization behavior: the ratio $\mathcal{C}(q)/g_0 q$ reaches very small values at the smallest accessible momenta, in full agreement with analytical expectations. This happens for those values of $\epsilon$, $T$, and $\gamma$, for which the localization length $\xi$ is smaller than the system size. For $\epsilon=0$ and $T=5$ [panel (a)], the strong localization is reached even for the smallest $\gamma/J$, which is in full consistency with the small value of $g_0$ for these parameters, see Fig.~\ref{Fig:Fig_1}.

We now inspect how this picture evolves when we move to 
panels (b) and (c), also corresponding to $\epsilon=0$ but to a larger period ($T=10$ and 20) and thus larger $g_0$. Now, for systems with a not too large measurement rate, the ratio remains rather far from reaching zero at our smallest momenta $q$. At the same time, the localization-induced non-monotonic behavior is perfectly observed. Going from large to small $q$, the curves first show the ballistic $1/q$ behavior at $q\ell_0 \gtrsim 1$ and then turn down in the diffusive range $q\ell_0 \ll 1$. Furthermore, the nearly straight-line form of the curves to the left of the maximum is perfectly consistent with the predicted logarithmic WL dependence, Eq.~\eqref{eq:Cq_over_q_weakloc}. 
   We now move to the panels (d)--(f), with the results for $\epsilon=3J/4$. For intermediate measurement strengths, with $\gamma/J > 0.1$, we observe the same non-monotonic behavior: the dependence of $\mathcal{C}(q)/g_0 q$ on $q$ shows a maximum and then turns down at lower $q$ as a manifestation of the WL correction. At the same time, for the smallest values of the measurement rate ($\gamma/J=0.02$, 0.05, and 0.08), the turning down is not observed.
   This is, however, fully expected and has a clear explanation. The momentum dependence of $\mathcal{C}(q)/g_0 q$ in the diffusive regime, $q\ell_0 \ll 1$, has two origins: a ballistic correction (a ``tail'' of the crossover from the ballistic to the diffusive regime), which decreases in magnitude with decreasing $q$, and a WL correction, whose magnitude increases with decreasing $q$. The position of the maximum results from a competition between these two effects. When $\gamma/J$ is decreased in panels (d)--(f), the values of $\ell_0$ and $g_0$ become correspondingly larger. The relative magnitude of the WL correction is proportional to $1/g_0$, so that one then needs smaller values of $q\ell_0$ to reach the maximum. Furthermore, the increase of $\ell_0$ makes this even more difficult. Thus, the curves with $\gamma/J=0.02$, 0.05, and 0.08 will turn down in the same way as curves with larger $\gamma$; this just requires smaller values of $q$, i.e., larger system sizes. 

   \subsubsection{Weak localization}
  \label{sec:numerics_Cq_weakloc}

\begin{figure}
\centering   \includegraphics[width=0.494\textwidth,height=0.24\textwidth]{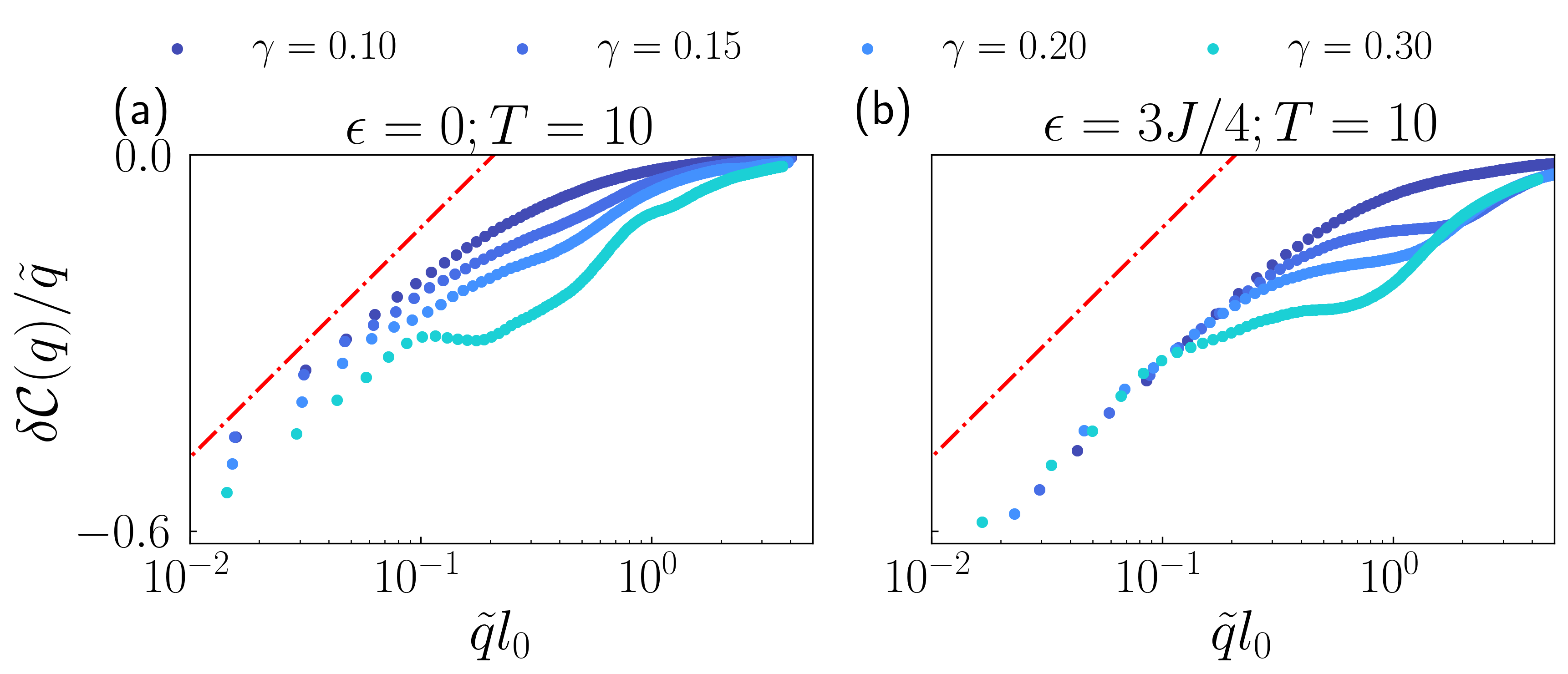}
    \caption{WL in density correlation function ${\cal C}(q)$. The plots show $\delta {\cal C}(q)/\tilde{q}$, where $\delta{\cal C}(q) = {\cal C}(q) - {\cal C}_0(q)$ is the difference between ${\cal C}(q)$ obtained from numerical simulations and ${\cal C}_0(q)$
that describes in the Gaussian approximation the ballistic-diffusive crossover  and is obtained from a solution of a Wiener-Hopf integral equation, see Sec.~\ref{Subsec:WH_integral}. Red dash-dotted lines have a slope corresponding to the analytically predicted WL correction in Eq.~\eqref{eq:Cq_over_q_weakloc}, following from the RG analysis of a NLSM of class BDI.
   }
    \label{Fig:Fig_7}
\end{figure}

   We now turn towards a quantitative analysis of the WL correction to the density correlation function.
   Ideally, one would like to focus on momenta much smaller than the position of the maximum, consider $\mathcal{C}(q)/q$ as a function of $\ln(q\ell_0)$ and compare with Eq.~\eqref{eq:Cq_over_q_weakloc}. However, for most of our data sets, the relevant range of $q$ is not so much smaller than the position of the maximum, which indicates that the ballistic correction is still sizable.
Thus, to improve the comparison, one should take into account the ballistic correction resulting from the ballistic-diffusive crossover studied  in Sec.~\ref{Subsec:WH_integral}.
This is done in Fig.~\ref{Fig:Fig_7} that shows $\delta {\cal C}(q)/\tilde{q}$, where $\delta{\cal C}(q) = {\cal C}(q) - {\cal C}_0(q)$ is the difference between ${\cal C}(q)$ obtained from numerical simulations and ${\cal C}_0(q)$
that describes the ballistic-diffusive crossover (neglecting quantum corrections) and is obtained from a numerical solution of a Wiener-Hopf integral equation, see Sec.~\ref{Subsec:WH_integral}.
The two panels of Fig.~\ref{Fig:Fig_7} show  $\delta {\cal C}(q)/\tilde{q}$
for $\epsilon=0$, $T=10$ and $\epsilon=3J/4$, $T=10$, respectively, both for several values of $\gamma/J$. It is seen that $\delta {\cal C}(q)$ is in a very good agreement with the analytically predicted WL correction, Eq.~\eqref{eq:Cq_over_q_weakloc} (shown by red dash-dotted lines), in the range of small momenta, $q\ell_0 \ll 1$.

Summarizing Sec.~\ref{Subsec:Density_correlation}, we conclude that the numerical data for the density correlation function results are in an excellent agreement with the analytical predictions using the mapping of the microscopic model to the NLSM of class BDI. In combination, these analytical results together with numerical analysis of the entanglement entropy and the density correlation function provide a rather complete and fully consistent understanding of the problem, both qualitative and quantitative. With increasing spatial scale $L$, the system crosses over from the ballistic regime to the diffusive (WL) one, and then to the strong-localization regime, with the characteristic crossover scales $\ell_0$ and $\xi$, respectively. There is thus no transition in the thermodynamic limit. At the same time, the localization length $\xi$ grows exponentially with the bare conductivity, $\xi \simeq \ell_0 e^{2\pi g_0}$. 

\section{Summary and perspectives}
\label{Sec:Conclusions}

In this paper, we have carried out a comprehensive analytical and numerical study of a periodically driven free-fermion system under continuous monitoring of the local particle density described by a stochastic Schr\"odinger equation. On the analytical side, we have extended the earlier developed mapping onto a NLSM to the case of periodically driven systems. We have also shown that the driving preserves the BDI symmetry class of the theory. The RG analysis of the NLSM leads to the conclusion that, for any values of microscopic parameters, the system exhibits the area-law behavior in the thermodynamic limit. When the system size $L$ (or, more generally, the spatial scale at which one probes physical properties) is increased, the system evolves through several consecutive regimes. For shortest scales, $L < \ell_0$, one finds the ballistic regime, with a volume-law growth of the entanglement entropy $S_{\rm vN}$. At intermediate scales $\ell_0 < L < \xi$, the system is in the diffusive (WL) regime, where $S_{\rm vN}$ exhibits a logarithmic growth with a negative WL correction. Finally, for $L>\xi$, we are in the strong-localization regime, where $S_{\rm vN}$ shows the area-law behavior (i.e., saturation). Here $\xi$ is the localization length that grows exponentially with the bare conductivity, $\xi \simeq \ell_0 \exp(2\pi g_0)$. 

While this behavior, which follows from the mapping to the NLSM, is universal, we have also studied important properties of the problem that depend on the microscopic model. This includes the behavior in the crossover from the ballistic to the diffusive regime. We have developed the corresponding analytical approach that leads to an integral equation of the Wiener-Hopf type. For generic momenta $q$, this equation can be solved numerically. In the small-$q$ limit, its analytical solution yields the key parameter of the NLSM theory: the bare conductivity $g_0$, 
see Eq.~\eqref{eq:g0} and Fig.~\ref{Fig:Fig_1}.

We have complemented the analytical study by numerical investigations of two quantities of central importance: the half-chain entanglement entropy $S_{\rm vN}(L)$ and the (momentum-space) density correlation function ${\cal C}(q)$. The numerical results  are in excellent agreement, both qualitative and quantitative, with the analytical predictions. This includes, first of all, the overall picture with ballistic, diffusive (WL), and strong-localization regimes,
which manifests itself in the data for $S_{\rm vN}(L)$, Fig.~\ref{Fig:Fig_2}, and, in
a particularly clear form, in results for ${\cal C}(q)$, 
Sec.~\ref{sec:numerics_Cq_overall} and Fig.~\ref{Fig:Fig_6}. Second, the slope of the logarithmic dependence of $S_{\rm vN}(L)$ in the initial part of the diffusive regime is in full consistency with the analytical prediction for the value of the bare conductivity $g_0$, see  Fig.~\ref{Fig:Fig_3}. Third, the analytical result for the WL correction with a universal prefactor corresponding to the BDI symmetry class is in full agreement with the numerical findings for $S_{\rm vN}(L)$
(Sec.~\ref{Subsubsec:WL}
and Fig.~\ref{Fig:Fig_4}) 
and ${\cal C}(q)$ (Sec.~\ref{sec:numerics_Cq_weakloc} and
Fig.~\ref{Fig:Fig_7}). Finally, our numerical results agree very well with the analytical prediction of the scaling of the saturation value of $S_{\rm vN}(L)$ in the strong-localization regime with the bare conductivity $g_0$, see
Sec.~\ref{Subsubsec:SL} and
Fig.~\ref{Fig:Fig_5}. 

It is worth emphasizing a crucial importance of the combination of analytical and numerical approaches in our work. If one would rely solely on numerical data, it would be very difficult to make a conclusion concerning the fate of the curves corresponding to large values of $g_0$, such as the data for $\gamma=0.02$ and 0.05 in  the panels (c) and (d) of Fig.~\ref{Fig:Fig_2}  and in the panels (d), (e), and (f) of
Fig.~\ref{Fig:Fig_6}. Indeed, by inspecting these numerical data, one could think that the behavior for these small values of $\gamma$ is qualitatively different from that for larger $\gamma$. This led the authors of Ref.~\cite{Modak_2025} to conjecture that, in the thermodynamic limit ($L\to \infty$) of this model, there is a BKT-type transition between the area-law phase and a logarithmic phase  at a certain finite measurement rate $\gamma_c$. Our analytical results, fully supported by numerical data, show that there is no such transition.  The curves with the smallest $\gamma$ in  the panels (c) and (d) of Fig.~\ref{Fig:Fig_2} will show the same bending down (towards saturation) as the curves with larger $\gamma$. Similarly, the curves with smallest $\gamma$ in panels (d), (e), and (f) of
Fig.~\ref{Fig:Fig_6} will exhibit a non-monotonic behavior towards $\mathcal{C}(q)/g_0 q \to 0$ at $q\to 0$, like the data for larger $\gamma$. However, this will happen at larger values of $L$ (correspondingly, smaller $q$) than those that were used in our simulations.

While the driving does not modify the universality class of the problem, it does enrich the physics of the problem in a major way. In particular, as we have shown, it strongly affects the value of the bare conductivity $g_0$, which in turn dramatically affects the value of the localization length $\xi$. This effect is particularly strong for symmetric drive, $\epsilon=0$, or for a weak asymmetry (small $\epsilon/J$). The parameters of the driving thus serve as additional knobs that allow one to move very efficiently between different regimes, which opens interesting perspectives for future research. In particular, in spatial dimensionality $d>1$ 
(or for Majorana fermions in $d=1$), one can let the system evolve through an MIPT by changing the drive period or asymmetry. Other promising avenues include incorporating interaction between fermions to study the interplay of periodic driving and monitoring in non-Gaussian systems and exploring a broader class of drive protocols, in particular, those that alter the universality class of the problem.

\begin{acknowledgments}
We thank Igor Gornyi for insightful discussions and feedback on this manuscript. We acknowledge support by the Deutsche Forschungsgemeinschaft (DFG, German Research Foundation) -- 553096561. The authors acknowledge support by the state of Baden-W{\"u}rttemberg through bwHPC, which was used for the numerical computations presented in this work.  
\end{acknowledgments}

\appendix

\section{Numerical simulation of SSE}
\label{app:SSE}

This appendix provides details of numerical simulation of the SSE given in Eq.~\eqref{Eq:SSE} of Sec.~\ref{Sec:Numerical_results}.

Any Gaussian state with the fixed number of particles $N$ is a Slater determinant, and can be obtained starting from the vacuum state $\ket{0}$ as 
\begin{equation}
\left| \Psi_t \right\rangle=\prod_{k=1}^{N}\left(\sum_{i=1}^{L}U_{i,k}(t)\,c_i^\dagger\right)|0\rangle,
\label{Eq:Evolution}
\end{equation}
where $U(t)$ is a matrix containing the filled single-particle orbitals. The orbitals can always be chosen to be orthonormal, implying that  $U^\dagger(t) U(t) = \mathbb{I}$. 
The complete description of the system and its evolution is then encoded in the correlation matrix
\begin{equation}
\mathcal{G}_{i,j}(t)=\left[U(t)U^\dagger(t)\right]_{j,i}=\left<\Psi_{t}\right|c^{\dagger}_{i}c_{j}\left|\Psi_{t}\right>,
\end{equation}
where the diagonal elements of this matrix determine the densities $\langle n_i \rangle_t$. 

Besides, the Trotterized evolution of the matrix $U$ for a single infinitesimal time step $dt$ follows from the SSE (Eq.~\eqref{Eq:SSE}), and is given as:
\begin{multline}
\label{eq:Ut}
U(t+dt)=\diag\left(\left\{ e^{d\xi_{i,t}+\gamma dt\left(2\left\langle n_{i}\right\rangle _{t}-1\right)}\right\} ^{L}_{i=1}\right)\\
\times e^{-i{\cal H}(t)dt}U(t),
\end{multline}
where $d\xi_{i, t}$ are independent and identically distributed Gaussian variables with variance $\overline{d\xi_{i,t}^2}=\gamma dt$. 
We explicitly conserve the orthonormality in the columns of $U$ using QR decomposition after each time step.

Throughout this work, we consider a lattice of size $L$ with periodic boundary conditions and start with a maximally disentangled N{\'e}el state $\left|\psi_{0}\right>=\left|1010\dots10\right>$, corresponding to half-filling $N = L/2$. 
We take $J=1$ as the unit of energy and consider the Trotterization time step $dt=0.05$ that yields sufficient accuracy. 
We then perform the evolution according to Eq.~\eqref{eq:Ut} up to sufficiently large time $t$ corresponding to the steady-state limit. 
We have verified that changing $dt$ or $t$ by a factor of  either $\frac12$ or two does not affect the results.  

For the numerical study of the half-chain entanglement entropy in Sec.~\ref{Subsec:Entanglement_entropy}, the system sizes in the range from $L=100$ to $L=1000$ were used. For investigation of the density correlation function in Sec.~\ref{Subsec:Density_correlation}, we used the system size $L=800$.

For each set of parameters, we have performed the averaging of studied quantities
over 50 quantum trajectories. In addition, we have averaged the half-chain entanglement entropy over the time and location of the subsystem cuts, see Sec.~\ref{Subsec:Entanglement_entropy} for detail.

\section{Ballistic-diffusive crossover}
\label{app:crossover}

In this appendix, we present details of the calculation of the density correlation function in the ballistic-diffusive crossover in the approximation that neglects loop corrections (Sec.~\ref{Subsec:WH_integral} of the main text).

For this calculation, we employ the diagrammatic technique based on the replica-Keldysh fermionic path integral and the perturbation theory with respect to the measurement-induced noise $\zeta(t)$. Evaluation of the density correlation function in the Gaussian approximation (that corresponds to neglecting loop corrections) amounts to a summation of diffuson-ladder diagrams.

In the steady-state limit, the information about the initial state  is lost (except for the conserved total charge). Thus, we can assume, without restricting generality,
 a spatially homogeneous initial state with the filling factor $n = 1/2$, which is the filling factor of the initial state as considered in the numerical analysis throughout the manuscript. 

The action (Eq.~\eqref{eq:FermionicLagrangian}) is restricted to the times at which the monitored evolution was performed. Assuming that we have started at $t_i \to -\infty$ and stopped at $t_f = 0$, all the time arguments in what follows are thus restricted to the domain $t < 0$.

It is easy to see that the average fermionic Green's functions, both bare and exact ones, are diagonal in the replica space. As a consequence, for their analysis it is sufficient to put $R = 1$ (which is equivalent to the analysis of the Lindbladian dynamics). For this reason, in this part of the calculation, we will omit the replica indices and focus only on the Keldysh structure of the Green's functions. 

In the presence of time-dependent Hamiltonian, the bare Green's function satisfy the following equation:
\begin{equation}
\left(i\partial_{t}-{\cal H}(t)\right)i\hat{G}^{(0)}(t,t^{\prime})=\delta(t-t^{\prime}),
\end{equation}
subject to the standard boundary conditions encoding the initial state \cite{KamenevLevchenko}. Since the Hamiltonian considered in the present manuscript is translation invariant at all times, the solution is particularly simple in the mixed time-momentum representation and is given by:
\begin{equation}
\label{eq:GGRGA0}
\hat{G}^{(0)}_{k}(t,t^{\prime})=G^{\text{R},(0)}_{k}(t,t^{\prime})\frac{1+\hat{\Lambda}}{2}+G^{\text{A},(0)}_{k}(t,t^{\prime})\frac{1-\hat{\Lambda}}{2},
\end{equation}
with the bare retarded and advanced Green's functions
\begin{align}
iG^{R,(0)}_{k}(t,t^{\prime})&=\Theta(t-t^{\prime})\exp\left(-i\int^{t}_{t^{\prime}}\varepsilon_{k}(\tau)d\tau\right),\\
iG^{A,(0)}_{k}(t,t^{\prime})&=-\Theta(t^{\prime}-t)\exp\left(i\int^{t^{\prime}}_{t}\varepsilon_{k}(\tau)d\tau\right),
\end{align}
and the Keldysh matrix encoding the initial state:
\begin{equation}
\hat{\Lambda}=\begin{pmatrix}1-2n & 2n\\
2(1-n) & -(1-2n)
\end{pmatrix}.
\end{equation}
At half-filling $n=1/2$ considered here, the matrix $\hat{\Lambda}$ reduces simply to $\hat{\Lambda} = \hat{\tau}_x$.

When considering the diagrammatic expansion with respect to $\zeta(t)$, one encounters a Green's function at coinciding temporal and spatial arguments. Unlike the standard diagrammatic technique convention, in the monitored setup, such diagrams are to be understood in the sense of principal values \cite{Poboiko_2023}. This element of the diagrammatic technique encodes the information about the average density only, which also holds true for the exact Green's functions. This allows us to write:
\begin{equation}
i\hat{{\cal G}}\equiv\int^{\pi}_{-\pi}\frac{dk}{2\pi}\frac{i}{2}\left(\hat{G}_{k}(t,t+0)+\hat{G}_{k}(t,t-0)\right)=\frac{1}{2}\hat{\Lambda},
\end{equation}

\begin{figure}
    \centering
    \includegraphics[width=\columnwidth]{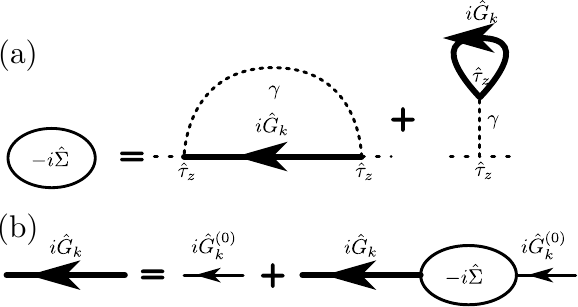}
    \caption{(a) The SCBA self-energy, as written in Eq.~\eqref{eq:SCBA}; and (b) the Dyson equation for the full Green's function as given in Eq.~\eqref{eq:DysonEq}.}
    \label{fig:SCBA}
\end{figure}

The next step is the self-consistent Born approximation (SCBA) for the Green's function. The self-energy is given by two the diagrams shown in Fig.~\ref{fig:SCBA}(a), which yields:
\begin{equation}
\label{eq:SCBA}
\hat{\Sigma}=-\gamma\left(\hat{\tau}_{z}\hat{{\cal G}}\hat{\tau}_{z}-\hat{\tau}_{z}\Tr\left(\hat{\tau}_{z}\hat{{\cal G}}\right)\right)=-\frac{i\gamma}{2}\hat{\Lambda}.
\end{equation}
We note that the SCBA for the Green's function is, in fact, exact \cite{Poboiko_2023}. 

The full Green's function is then determined from the Dyson equation depicted graphically in Fig.~\ref{fig:SCBA}(b) as:
\begin{equation}
\label{eq:DysonEq}
\left(i\partial_{t}-{\cal H}(t)-\hat{\Sigma}\right)i\hat{G}=\hat{\mathbb{I}}.
\end{equation}
For the full Green's functions, this yields  the structure analogous to \eqref{eq:GGRGA0} given by,
\begin{equation}
\label{eq:GGRGA}
\hat{G}_{k}(t,t^{\prime})=G^{\text{R}}_{k}(t,t^{\prime})\frac{1+\hat{\Lambda}}{2}+G^{\text{A}}_{k}(t,t^{\prime})\frac{1-\hat{\Lambda}}{2},
\end{equation}
with 
\begin{equation}
G^{\text{R}/\text{A}}_{k}(t,t^{\prime})=G^{\text{R}/\text{A},(0)}_{k}(t,t^{\prime})\exp\left(-\frac{\gamma}{2}\left|t-t^{\prime}\right|\right),
\end{equation}
which is exactly Eq.~\eqref{eq:GR} of the main text.

\begin{figure}
    \centering
    \includegraphics[width=\columnwidth]{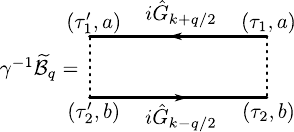}
    \caption{Diffuson block $\check{{\cal B}}_q$ given by Eq.~\eqref{eq:Bq:Tensor}. Here, $\tau_{1,2}$ and $\tau_{1,2}^\prime$ are Keldysh indices, while $a, b$ denote replica indices.}
    \label{fig:block}
\end{figure}

The key building block of the diagrammatic evaluation of the diffuson ladder is the ``diffuson block'', which is a loop diagram consisting of two Green's functions, illustrated in Fig.~\ref{fig:block}. In agreement with the standard diagramatic rules, we attribute an extra ``minus'' sign to the fermionic loop (which compensates for $i^2$ arising from two Green's functions). Utilizing the representation in Eq.~\eqref{eq:GGRGA}, we then obtain:
\begin{align}
\label{eq:Bq:Tensor}
\widetilde{{\cal B}}_{q}(t,t^{\prime})\equiv \gamma \int^{\pi}_{-\pi}\frac{dk}{2\pi}\hat{G}_{k+q/2}(t,t^{\prime})\otimes\hat{G}^{T}_{k-q/2}(t^{\prime},t)\\
={\cal B}_{q}(t,t^{\prime})\widetilde{\mathbb{P}}^{R}+{\cal B}_{-q}(t^{\prime},t)\widetilde{\mathbb{P}}^{A},
\end{align}
with the scalar block given by Eq.~\eqref{eq:WH:Bq} of the main text, and with the tensor Keldysh structure encoded in mutually orthogonal projectors:
\begin{equation}
\hat{\mathbb{P}}^{R}_{\tau_{1}\tau_{2},\tau^{\prime}_{1}\tau^{\prime}_{2}}=\left(\frac{1+\hat{\Lambda}}{2}\right)_{\tau_{1}\tau^{\prime}_{1}}\otimes\left(\frac{1-\hat{\Lambda}^{T}}{2}\right)_{\tau_{2}\tau^{\prime}_{2}}=\hat{\mathbb{P}}^{A}_{\tau^{\prime}_{2}\tau^{\prime}_{1},\tau_{2}\tau_{1}}.
\end{equation}
Here Keldysh indices $(\tau_1,\tau_1^\prime)$ and $(\tau_2,\tau_2^\prime)$ correspond to the upper and lower Green's functions, respectively, as demonstrated in Fig.~\ref{fig:block}. 

Then, the $4 \times 4$ structure of the block can be conveniently represented as:
\begin{equation}
\label{eq:Bstructure}
\widetilde{{\cal B}}_{q}=U\check{{\cal B}}_{q}V^{T},
\end{equation}
with
\begin{equation}
\check{{\cal B}}_{q}=\begin{pmatrix}{\cal B}_{q} & 0\\
0 & {\cal B}^{T}_{q}
\end{pmatrix},\quad U=V=\frac{1}{2}\begin{pmatrix}1 & 1\\
-1 & 1\\
1 & -1\\
-1 & -1
\end{pmatrix},
\end{equation}
satisfying $V^T U = 1$ (we note that, while the matrices $U$ and $V$ are equal only for $n = 1/2$, such a representation can be written also for arbitrary $n$). In this form, we explicitly separate the $2 \times 2$ ``retarded-advanced'' sector. We define ${\cal B}_q^T(t,t^\prime) \equiv {\cal B}_{-q}(t^\prime, t)$.

\begin{figure}
    \centering
    \includegraphics[width=\columnwidth]{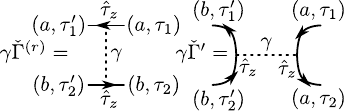}
    \caption{Diffuson vertices, as given in Eqs.~(\ref{eq:V1}) and (\ref{eq:V2}).}
    \label{fig:vertices}
\end{figure}

Now, we proceed to the calculation of the diffuson ladder. The measurement-induced noise yields two ladder vertices depicted in Fig.~\ref{fig:vertices} given by:
\begin{align}
\label{eq:V1}
\widetilde{\Gamma}^{(r)}_{\tau_{1}\tau_{2},\tau_{1}^\prime \tau_2^\prime}&=-\tau^{z}_{\tau_{1}\tau_{1}^\prime}\tau^{z}_{\tau_2 \tau_2^\prime},\\
\label{eq:V2}
\widetilde{\Gamma}^{\prime}_{\tau_{1}\tau_{2},\tau_{1}^\prime \tau_2^\prime}&=\tau^{z}_{\tau_{1}\tau_2}\tau^{z}_{\tau_{1}^\prime \tau_2^\prime},
\end{align}
which have distinct Keldysh and replica structures. The additional minus sign for $\check{\Gamma}^{(r)}$ vertex accounts for the fact that this diagram does not add an extra fermionic loop, thus we need an extra minus sign to compensate for the minus sign attributed to the definition of $\widetilde{{\cal B}}_q$.

\begin{figure}
    \centering
    \includegraphics[width=\columnwidth]{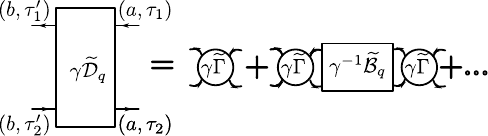}
    \caption{Diffuson ladder corresponding to the solution of Eq.~\eqref{eq:DiffusonEq}.}
    \label{fig:diffuson_ladder}
\end{figure}

Generically, the two-body Green's function depends on four replica and Keldysh indices, corresponding to four fermionic legs in the diagram. For our purposes, however, it is sufficient to consider only the ladder with equal rightmost replica indices and equal leftmost replica indices (see
Fig.~\ref{fig:diffuson_ladder} where these indices are denoted by $a$ and $b$, respectively).
The diffuson then becomes a $R \times R$ matrix in the replica space and a $4 \times 4$ tensor in the Keldysh space.
Furthermore, the vertex $\Gamma^{(r)}$ does not change the replica indices of the Green's functions, and is thus a diagonal matrix in the replica space, whereas $\Gamma^\prime$ allows for arbitrary change of the replica index. The combined vertex thus reads:
\begin{equation}
\widetilde{\Gamma}_{ab}=\widetilde{\Gamma}^{(r)}\delta_{ab}+\left(\widetilde{\Gamma}^{(s)}-\widetilde{\Gamma}^{(r)}\right){\cal I}_{ab},
\end{equation}
with the replica-mixing matrix ${\cal I}$ having all matrix elements equal to unity, ${\cal I}_{a b} = 1$, and
\begin{equation}
    \widetilde{\Gamma}^{(s)} = \widetilde{\Gamma}^{(r)} + \widetilde{\Gamma}^\prime \,.
\end{equation}

The diffuson then satisfies the following equation:
\begin{equation}
\label{eq:DiffusonEq}
\left(1-\widetilde{\Gamma} \widetilde{{\cal B}}_{q}\right)\widetilde{{\cal {\cal D}}}_{q}=\widetilde{\Gamma}.
\end{equation}
Furthermore, the replica structure of the solution is inherited from that of the vertex, obtained as:
\begin{equation}
\widetilde{{\cal D}}_{q,ab}=\widetilde{{\cal D}}^{(r)}_{q}\delta_{ab}+\left(\widetilde{{\cal D}}^{(s)}_{q}-\widetilde{{\cal D}}^{(r)}_{q}\right){\cal I}_{ab},
\end{equation}
where 
\begin{equation}
\widetilde{{\cal D}}^{(r/s)}_{q}=\left(1- \widetilde{\Gamma}^{(r/s)}\widetilde{{\cal B}}^{(r/s)}_{q}\right)^{-1}\hat{\Gamma}^{(r/s)}.
\end{equation}
These $4 \times 4$ matrices can be inverted analytically owing to the block structure as given in \eqref{eq:Bstructure}, and is written as:
\begin{equation}
\widetilde{{\cal D}}^{(s/r)}_q= \Gamma^{(s/r)}+\Gamma^{(s/r)}U\check{{\cal D}}^{(s/r)}V^{T}\Gamma^{(s/r)},
\end{equation}
with the following $2 \times 2$ matrices describing the replica-symmetric and the replicon diffusons:
\begin{equation}
\check{{\cal D}}^{(s)}_{q}=\begin{pmatrix}\left[1-{\cal B}_{q}\right]^{-1}{\cal B}_{q} & 0\\
0 & \left[1-{\cal B}^{T}_{q}\right]^{-1}{\cal B}^{T}_{q}
\end{pmatrix},
\end{equation}
\begin{equation}
\check{{\cal D}}^{(r)}_{q}=\begin{pmatrix}{\cal D}_{q}{\cal B}_{q} & 1-{\cal D}_{q}\\[6pt]
-{\cal B}^{T}_{q}{\cal D}_{q}{\cal B}_{q} & {\cal B}^{T}_{q}{\cal D}_{q}
\end{pmatrix},
\end{equation}
where the scalar diffuson ${\cal D}_q$ of the replicon sector is given by Eq.~\eqref{eq:integral-operator-Dq} of the main text.

Finally, we proceed to the calculation of the two-body Green's function, which is given by the diffuson with additional blocks attached:
\begin{equation}
\widetilde{{\cal C}}_{q}=\gamma^{-1}\left(\widetilde{{\cal B}}_{q}+\widetilde{{\cal B}}_{q}\widetilde{{\cal D}}_{q}\widetilde{{\cal B}}_{q}\right)=\widetilde{{\cal C}}^{(r)}_{q}\delta_{ab}+\left({\cal \widetilde{C}}^{(s)}_{q}-\widetilde{{\cal C}}^{(r)}_{q}\right){\cal I}_{ab},
\end{equation}
where
\begin{equation}
\widetilde{{\cal C}}^{(s,r)}_{q}=\gamma^{-1}U{\cal D}^{(s/r)}_{q}V^{T}\,.
\end{equation}
This gives rise to two physically distinct correlation functions, replica-symmetric and replicon ones:
\begin{align}
{\cal C}^{(s)}(\boldsymbol{x}-\boldsymbol{y})&\equiv\overline{\left\langle \hat{n}_{\boldsymbol{x}}\hat{n}_{\boldsymbol{y}}\right\rangle }-\overline{\left\langle \hat{n}_{\boldsymbol{x}}\right\rangle }\,\,\overline{\left\langle \hat{n}_{\boldsymbol{y}}\right\rangle }\\
{\cal C}^{(r)}(\boldsymbol{x}-\boldsymbol{y})&\equiv\overline{\left\langle \hat{n}_{\boldsymbol{x}}\hat{n}_{\boldsymbol{y}}\right\rangle }-\overline{\left\langle \hat{n}_{\boldsymbol{x}}\right\rangle \left\langle \hat{n}_{\boldsymbol{y}}\right\rangle },
\end{align}
where bold symbols $\boldsymbol{x} = (x,t), \boldsymbol{y} = (y,t^\prime)$ denote combined temporal and spatial arguments. These physically relevant density correlation functions are 
obtained from the full two-body Green's function by taking the limit $t, t^\prime \to -0$ and attaching ``classical'' Keldysh vertices $\hat{\tau}^z/ 2$ to both ends as:
\begin{equation}
{\cal C}^{(r/s)}_{q}=\frac{1}{4 \gamma}\tau^{z}_{\tau_{1}\tau_{2}}\widetilde{{\cal C}}^{(r/s)}_{q,\tau_{1}\tau_{2},\tau^{\prime}_{1}\tau^{\prime}_{2}}\tau^{z}_{\tau^{\prime}_{1}\tau^{\prime}_{2}}=\frac{1}{4 \gamma}\left<\tau_{z}\right|\widetilde{{\cal C}}_{q}\left|\tau_{z}\right>.
\end{equation}
For the replicon sector $\mathcal{C}_q \equiv \mathcal{C}_q^{(r)}$, which yields Eq.~\eqref{eq:Cq-operator} of the main text.

\section{Diffusion constant}
\label{app:g0}

This appendix provides details on the derivation of Eq.~\eqref{eq:g0} for the dimensionless coupling constant $g_0$, which enters the NLSM action in Eq.~\eqref{eq:NLSMAction} and
plays the role of the effective 2D diffusion constant. 

The value of $g_0$ can be conveniently obtained by means of the long-wavelength expansion of the ballistic-diffusive crossover behavior described in Appendix~\ref{app:crossover}, which is equivalent to the gradient expansion one performs in the standard derivation of the NLSM. 
Furthermore, the diffusion constant is a property of the bulk, both spatial and temporal. This implies that, unlike the density-density correlation function for which the presence of the boundary in the time domain at $t = t_f = 0$ is crucial, the diffusion constant can be  safely evaluated by putting $t_f \to +\infty$.

At $q \to 0$, the diffuson block as given in Eq.~\eqref{eq:WH:Bq} is expanded as:
\begin{align}
{\cal B}_{q}(t,t^{\prime})&\approx A(t,t^{\prime})-q^{2}B(t,t^{\prime});\\
A(t,t^{\prime})&=\tau^{-1}_{0}\Theta(t-t^{\prime})\exp\left(-\frac{t-t^{\prime}}{\tau_{0}}\right);\\
B(t,t^{\prime})&=\tau^{-1}_{0}\Theta(t-t^{\prime})\exp\left(-\frac{t-t^{\prime}}{\tau_{0}}\right)\left(\int^{t}_{t^{\prime}}J(\tau)d\tau\right)^{2}.
\end{align}
The operator $A$ is translationally invariant in time, and its Fourier transform is readily found to be
\begin{equation}
A(\omega)=\frac{1}{1-i\omega\tau_{0}}\approx1+i\omega\tau_{0}-\frac{1}{2}\omega^{2}\tau^{2}_{0}.
\end{equation}

For the long-wavelength expansion of the diffusion propagator as written in Eq.~\eqref{eq:integral-operator-Dq} of the main text, we obtain:
\begin{align}
{\cal D}_{q}(t,t^{\prime})&=\left[1-\hat{{\cal B}}_{q}\hat{{\cal B}}^{T}_{q}\right]^{-1}\notag\\
&\approx\left[1-\hat{A}\hat{A}^{T}+q^{2}\left(\hat{B}\hat{A}^{T}+\hat{B}^{T}\hat{A}\right)\right]^{-1}\notag\\
&\approx\frac{1}{4\tau_0}\left[D_t\omega^{2}+\hat{D}_xq^{2}\right]^{-1}.
\end{align}
Here, the temporal diffusion constant and spatial diffusion operator are given by
\begin{equation}
\label{eq:appendix-Dt}
D_t = \frac{\tau_0}{4},\quad \hat{D}_x = \frac{1}{4\tau_0}\left(\hat{B}\hat{A}^{T}+\hat{B}^{T}\hat{A}\right)\approx \frac{1}{4\tau_0}\left(\hat{B}+\hat{B}^{T}\right),
\end{equation}
with the kernel
\begin{equation}
D_x(t,t^{\prime})\approx \frac{1}{4\tau_0^2}\exp\left(-\frac{|t-t^{\prime}|}{\tau_{0}}\right)\left(\int^{t}_{t^{\prime}}J(\tau)d\tau\right)^{2}.
\end{equation}
Note that the operator $\hat{D}_x$ is a Floquet operator, i.e. its kernel satisfies $D_x(t+T,t^\prime+T) = D_x(t,t^\prime)$, thus its eigenvalues and eigenvectors can be characterized by quasi-frequency $\omega \in [-\pi/T, \pi/T)$. 
Furthermore, since it is multiplied by $q^2$, it is sufficient to evaluate it at zero quasi-frequency $\omega = 0$; in the same limit, one has $A(\omega=0) = 1$. 

The characteristic values of momentum $q$ are determined by the inverse system size $q \sim L^{-1}$ (more generally, the role of $L$ here is played by the spatial scale at which the problem is studied).
When these values of $q$ are small compared to the scale set by the inverse characteristic Floquet bandwidth, i.e., $D_t / T^2 \gg D_x / L^2$ or, equivalently, $L \gg T \sqrt{D_t / D_x}$, the behavior of $\mathcal{D}_q(t,t^\prime)$ is dominated by the lowest Floquet band.
Formally, this means that the macroscopic diffusion constant can be calculated by performing a Wigner transformation of the operator $\hat{D}_x$ and considering only its zero Fourier harmonic with respect to the center-of-mass time $(t + t^\prime) / 2$:
\begin{equation}
D_{x}=\int^{T}_{0}\frac{dt_{1}}{T}\int^{\infty}_{-\infty}d\tau\,D_{x}\left(t_{1}+\frac{\tau}{2},t_{1}-\frac{\tau}{2}\right).
\end{equation}
For the drive considered in the present manuscript, i.e., $J(t) = J_1$ for half-period and $J_2$ for another half-period, the integral can be easily calculated and yields:
\begin{equation}
\label{eq:appendix-Dx}
D_x = \frac{\tau_{0}}{4}\left(\left(J_{1}+J_{2}\right)^{2}+\left(J_{1}-J_{2}\right)^{2}{\cal I}\left(\frac{T}{4\tau_{0}}\right)\right),
\end{equation}
with the function
\begin{equation}
{\cal I}(p)\equiv1-\frac{\tanh p}{p}.
\end{equation}

To establish a correspondence between the diffusion constants $D_x$, Eq.~\eqref{eq:appendix-Dx} and $D_t$, 
Eq.~\eqref{eq:appendix-Dt}, on one hand, and the couplings of the NLSM, on the other hand, we consider the \emph{bulk} density-density correlation function 
(i.e., the one neglecting the presence of the boundary) in both formalisms. The microscopic diagrammatic calculation yields Eq.~\eqref{eq:Cq-operator}:
\begin{align}
\label{eq:appendix-C_q_omega}
{\cal C}(q,\omega)&\approx\frac{\tau_{0}}{4}\left[1-\left(1-A(\omega)\right){\cal D}(q,\omega)\left(1-A^{T}(\omega)\right)\right]\notag\\
&=\frac{D_{x}D_{t}q^{2}}{D_{t}\omega^{2}+D_{x}q^{2}}.
\end{align}
On the other hand, it has been shown (refer to Ref.~\cite{Poboiko_2026}) that, for the NLSM action of the form
\begin{equation}
\label{eq:appendix-NLSM-action}
{\cal L}[U]=\frac{1}{4}\Tr\left(D_{t}\,\partial_{t}U^{\dagger}\partial_{t}U+D_{x}\,\partial_{x}U^{\dagger}\partial_{x}U\right),
\end{equation}
the ``bulk'' density-density correlation function, which corresponds to the correlation function of the NLSM N{\"o}ther currents ${\cal J}=-iU^{\dagger}\partial_{t}U$ (with $\boldsymbol{r} = (x,t)$),
\begin{equation}
{\cal C}(\boldsymbol{r}-\boldsymbol{r}^{\prime})\simeq D_{t}\delta(\boldsymbol{r}-\boldsymbol{r}^{\prime})-D^{2}_{t}\left\langle {\cal J}_{ab}(\boldsymbol{r}){\cal J}_{ba}(\boldsymbol{r}^{\prime})\right\rangle,
\end{equation}
has exactly the same form as obtained in Eq.~\eqref{eq:appendix-C_q_omega}. (Here $a \neq b$ are two different combined indices in the replica and the particle-hole space, see Ref.~\cite{Poboiko_2026}.)
This allows us to identify $D_t$ and $D_x$ which enter the diffuson ladder in Eqs.~\eqref{eq:appendix-Dt} and \eqref{eq:appendix-Dx},
with the coefficients $D_t$ and $D_x$ in the NLSM action (Eq.~\eqref{eq:appendix-NLSM-action}).
A rescaling of the coordinates then allows one to bring the NLSM action to the isotropic form, as in Eq.~\eqref{eq:NLSMAction}, with the coupling constant that is obtained as the geometric mean of both diffusion constants:
\begin{equation}
g_{0}=\sqrt{D_x D_t}.
\label{eq:app-g0}
\end{equation}
Upon substituting Eq.~\eqref{eq:appendix-Dt} for $D_t$ and Eq.~\eqref{eq:appendix-Dx} for $D_x$,
we obtain the expression for the bare diffusion constant as presented in Eq.~\eqref{eq:g0} of the main text.

The diffusion constant $D_x$ and the mean free time $\tau_0$ allow us to estimate the UV scale (mean free path) $\ell_0$ via $\ell_0^2 = 2 D_{x} \tau_0$. In combination with Eqs.~\eqref{eq:app-g0} and \eqref{eq:appendix-Dt}, this yields the relation $\ell_0 = 2\sqrt{2}g_0$. In the absence of driving, this formula was obtained in Ref.~\cite{Poboiko_2023} (up to a minor difference in a factor $\sqrt{2}$, which is due to a difference between SSE and projective measurements).

\bibliography{references,refs}

@article{Li_2018,
  title = {Quantum Zeno effect and the many-body entanglement transition},
  author = {Li, Yaodong and Chen, Xiao and Fisher, Matthew P. A.},
  journal = {Phys. Rev. B},
  volume = {98},
  issue = {20},
  pages = {205136},
  numpages = {9},
  year = {2018},
  month = {Nov},
  publisher = {American Physical Society},
  doi = {10.1103/PhysRevB.98.205136},
  url = {https://link.aps.org/doi/10.1103/PhysRevB.98.205136}
}

@article{Li_2019,
  title = {Measurement-driven entanglement transition in hybrid quantum circuits},
  author = {Li, Yaodong and Chen, Xiao and Fisher, Matthew P. A.},
  journal = {Phys. Rev. B},
  volume = {100},
  issue = {13},
  pages = {134306},
  numpages = {26},
  year = {2019},
  month = {Oct},
  publisher = {American Physical Society},
  doi = {10.1103/PhysRevB.100.134306},
  url = {https://link.aps.org/doi/10.1103/PhysRevB.100.134306}
}

@article{Smith_2019,
  title = {Unitary-projective entanglement dynamics},
  author = {Chan, Amos and Nandkishore, Rahul M. and Pretko, Michael and Smith, Graeme},
  journal = {Phys. Rev. B},
  volume = {99},
  issue = {22},
  pages = {224307},
  numpages = {16},
  year = {2019},
  month = {Jun},
  publisher = {American Physical Society},
  doi = {10.1103/PhysRevB.99.224307},
  url = {https://link.aps.org/doi/10.1103/PhysRevB.99.224307}
}

@article{Skinner_2019,
  title = {Measurement-Induced Phase Transitions in the Dynamics of Entanglement},
  author = {Skinner, Brian and Ruhman, Jonathan and Nahum, Adam},
  journal = {Phys. Rev. X},
  volume = {9},
  issue = {3},
  pages = {031009},
  numpages = {21},
  year = {2019},
  month = {Jul},
  publisher = {American Physical Society},
  doi = {10.1103/PhysRevX.9.031009},
  url = {https://link.aps.org/doi/10.1103/PhysRevX.9.031009}
}

@article{Cao_2019,
	title = {Entanglement in a fermion chain under continuous monitoring},
	pages = {024},
	author = {Cao, Xiangyu and Tilloy, Antoine and De Luca, Andrea},
	journal = {SciPost Phys.},
	volume = {7},
	year = {2019},
	publisher = {SciPost},
	doi = {10.21468/SciPostPhys.7.2.024},
	url = {https://scipost.org/10.21468/SciPostPhys.7.2.024}
}

@article{Vijay_2012,
	title = {Stabilizing Rabi oscillations in a superconducting qubit using quantum feedback},
	pages = {77-80},
	author = {Vijay, R. and Macklin, C. and Slichter, D. H. and Weber, S. J. and Murch, K. W. and Naik, R. and Korotkov, A. N. and Siddiqi, I.},
	journal = {Nature},
	volume = {490},
	year = {2012},
	publisher = {Nature},
	doi = {10.1038/nature11505},
	url = {https://doi.org/10.1038/nature11505}
}

@article{Katz_2006,
author = {N. Katz  and M. Ansmann  and Radoslaw C. Bialczak  and Erik Lucero  and R. McDermott  and Matthew Neeley  and Matthias Steffen  and E. M. Weig  and A. N. Cleland  and John M. Martinis  and A. N. Korotkov },
title = {Coherent State Evolution in a Superconducting Qubit from Partial-Collapse Measurement},
journal = {Science},
volume = {312},
number = {5779},
pages = {1498-1500},
year = {2006},
doi = {10.1126/science.1126475},
URL = {https://www.science.org/doi/abs/10.1126/science.1126475},
eprint = {https://www.science.org/doi/pdf/10.1126/science.1126475}
}

@article{Huard_2016,
  title = {Observing Quantum State Diffusion by Heterodyne Detection of Fluorescence},
  author = {Campagne-Ibarcq, P. and Six, P. and Bretheau, L. and Sarlette, A. and Mirrahimi, M. and Rouchon, P. and Huard, B.},
  journal = {Phys. Rev. X},
  volume = {6},
  issue = {1},
  pages = {011002},
  numpages = {7},
  year = {2016},
  month = {Jan},
  publisher = {American Physical Society},
  doi = {10.1103/PhysRevX.6.011002},
  url = {https://link.aps.org/doi/10.1103/PhysRevX.6.011002}
}

@article{Buchhold_2021,
  title = {Effective Theory for the Measurement-Induced Phase Transition of Dirac Fermions},
  author = {Buchhold, M. and Minoguchi, Y. and Altland, A. and Diehl, S.},
  journal = {Phys. Rev. X},
  volume = {11},
  issue = {4},
  pages = {041004},
  numpages = {35},
  year = {2021},
  month = {Oct},
  publisher = {American Physical Society},
  doi = {10.1103/PhysRevX.11.041004},
  url = {https://link.aps.org/doi/10.1103/PhysRevX.11.041004}
}

@article{Modak_2025,
  title = {Measurement-induced phase transition in periodically driven free-fermionic systems},
  author = {Chatterjee, Pallabi and Modak, Ranjan},
  journal = {Phys. Rev. B},
  volume = {112},
  issue = {2},
  pages = {024304},
  numpages = {19},
  year = {2025},
  month = {Jul},
  publisher = {American Physical Society},
  doi = {10.1103/8l26-c7cv},
  url = {https://link.aps.org/doi/10.1103/8l26-c7cv}
}

@article{Szyniszewski_2019,
  title = {Entanglement transition from variable-strength weak measurements},
  author = {Szyniszewski, M. and Romito, A. and Schomerus, H.},
  journal = {Phys. Rev. B},
  volume = {100},
  issue = {6},
  pages = {064204},
  numpages = {8},
  year = {2019},
  month = {Aug},
  publisher = {American Physical Society},
  doi = {10.1103/PhysRevB.100.064204},
  url = {https://link.aps.org/doi/10.1103/PhysRevB.100.064204}
}

@article{Szyniszewski_2023,
  title = {Disordered monitored free fermions},
  author = {Szyniszewski, Marcin and Lunt, Oliver and Pal, Arijeet},
  journal = {Phys. Rev. B},
  volume = {108},
  issue = {16},
  pages = {165126},
  numpages = {12},
  year = {2023},
  month = {Oct},
  publisher = {American Physical Society},
  doi = {10.1103/PhysRevB.108.165126},
  url = {https://link.aps.org/doi/10.1103/PhysRevB.108.165126}
}

@article{Szyniszewski_2020,
  title = {Universality of Entanglement Transitions from Stroboscopic to Continuous Measurements},
  author = {Szyniszewski, M. and Romito, A. and Schomerus, H.},
  journal = {Phys. Rev. Lett.},
  volume = {125},
  issue = {21},
  pages = {210602},
  numpages = {6},
  year = {2020},
  month = {Nov},
  publisher = {American Physical Society},
  doi = {10.1103/PhysRevLett.125.210602},
  url = {https://link.aps.org/doi/10.1103/PhysRevLett.125.210602}
}

@article{Yang_2018,
  title = {Theory of a Quantum Scanning Microscope for Cold Atoms},
  author = {Yang, D. and Laflamme, C. and Vasilyev, D. V. and Baranov, M. A. and Zoller, P.},
  journal = {Phys. Rev. Lett.},
  volume = {120},
  issue = {13},
  pages = {133601},
  numpages = {6},
  year = {2018},
  month = {Mar},
  publisher = {American Physical Society},
  doi = {10.1103/PhysRevLett.120.133601},
  url = {https://link.aps.org/doi/10.1103/PhysRevLett.120.133601}
}

@article{Alberton_2021,
  title = {Entanglement Transition in a Monitored Free-Fermion Chain: From Extended Criticality to Area Law},
  author = {Alberton, O. and Buchhold, M. and Diehl, S.},
  journal = {Phys. Rev. Lett.},
  volume = {126},
  issue = {17},
  pages = {170602},
  numpages = {6},
  year = {2021},
  month = {Apr},
  publisher = {American Physical Society},
  doi = {10.1103/PhysRevLett.126.170602},
  url = {https://link.aps.org/doi/10.1103/PhysRevLett.126.170602}
}

@article{Gisin_1992,
doi = {10.1088/0305-4470/25/21/023},
url = {https://doi.org/10.1088/0305-4470/25/21/023},
year = {1992},
month = {nov},
publisher = {},
volume = {25},
number = {21},
pages = {5677},
author = {N Gisin and I C Percival},
title = {The quantum-state diffusion model applied to open systems},
journal = {Journal of Physics A: Mathematical and General}
}

@article{Alonso_2017,
  title = {Dynamics of non-Markovian open quantum systems},
  author = {de Vega, In\'es and Alonso, Daniel},
  journal = {Rev. Mod. Phys.},
  volume = {89},
  issue = {1},
  pages = {015001},
  numpages = {58},
  year = {2017},
  month = {Jan},
  publisher = {American Physical Society},
  doi = {10.1103/RevModPhys.89.015001},
  url = {https://link.aps.org/doi/10.1103/RevModPhys.89.015001}
}

@article{Gisin_1998,
  title = {Non-Markovian quantum state diffusion},
  author = {Di\'osi, L. and Gisin, N. and Strunz, W. T.},
  journal = {Phys. Rev. A},
  volume = {58},
  issue = {3},
  pages = {1699--1712},
  numpages = {0},
  year = {1998},
  month = {Sep},
  publisher = {American Physical Society},
  doi = {10.1103/PhysRevA.58.1699},
  url = {https://link.aps.org/doi/10.1103/PhysRevA.58.1699}
}

@article{Calabrese_2005,
doi = {10.1088/1742-5468/2005/04/P04010},
url = {https://doi.org/10.1088/1742-5468/2005/04/P04010},
year = {2005},
month = {apr},
publisher = {},
volume = {2005},
number = {04},
pages = {P04010},
author = {Calabrese, Pasquale and Cardy, John},
title = {Evolution of entanglement entropy in one-dimensional systems},
journal = {Journal of Statistical Mechanics: Theory and Experiment}
}

@article{Agrawal_2022,
  title = {Entanglement and Charge-Sharpening Transitions in U(1) Symmetric Monitored Quantum Circuits},
  author = {Agrawal, Utkarsh and Zabalo, Aidan and Chen, Kun and Wilson, Justin H. and Potter, Andrew C. and Pixley, J. H. and Gopalakrishnan, Sarang and Vasseur, Romain},
  journal = {Phys. Rev. X},
  volume = {12},
  issue = {4},
  pages = {041002},
  numpages = {29},
  year = {2022},
  month = {Oct},
  publisher = {American Physical Society},
  doi = {10.1103/PhysRevX.12.041002},
  url = {https://link.aps.org/doi/10.1103/PhysRevX.12.041002}
}

@article{Jacobs_2006,
author = {Kurt Jacobs and Daniel A. Steck},
title = {A straightforward introduction to continuous quantum measurement},
journal = {Contemporary Physics},
volume = {47},
number = {5},
pages = {279--303},
year = {2006},
publisher = {Taylor \& Francis},
doi = {10.1080/00107510601101934},
URL = {https://doi.org/10.1080/00107510601101934}
}

@article{Wiseman_1993,
  title = {Quantum theory of field-quadrature measurements},
  author = {Wiseman, H. M. and Milburn, G. J.},
  journal = {Phys. Rev. A},
  volume = {47},
  issue = {1},
  pages = {642--662},
  numpages = {0},
  year = {1993},
  month = {Jan},
  publisher = {American Physical Society},
  doi = {10.1103/PhysRevA.47.642},
  url = {https://link.aps.org/doi/10.1103/PhysRevA.47.642}
}

@article{Jian_2020,
  title = {Measurement-induced criticality in random quantum circuits},
  author = {Jian, Chao-Ming and You, Yi-Zhuang and Vasseur, Romain and Ludwig, Andreas W. W.},
  journal = {Phys. Rev. B},
  volume = {101},
  issue = {10},
  pages = {104302},
  numpages = {11},
  year = {2020},
  month = {Mar},
  publisher = {American Physical Society},
  doi = {10.1103/PhysRevB.101.104302},
  url = {https://link.aps.org/doi/10.1103/PhysRevB.101.104302}
}

@article{Bao_2020,
  title = {Theory of the phase transition in random unitary circuits with measurements},
  author = {Bao, Yimu and Choi, Soonwon and Altman, Ehud},
  journal = {Phys. Rev. B},
  volume = {101},
  issue = {10},
  pages = {104301},
  numpages = {26},
  year = {2020},
  month = {Mar},
  publisher = {American Physical Society},
  doi = {10.1103/PhysRevB.101.104301},
  url = {https://link.aps.org/doi/10.1103/PhysRevB.101.104301}
}

@article{Gullans_2020,
  title = {Dynamical Purification Phase Transition Induced by Quantum Measurements},
  author = {Gullans, Michael J. and Huse, David A.},
  journal = {Phys. Rev. X},
  volume = {10},
  issue = {4},
  pages = {041020},
  numpages = {28},
  year = {2020},
  month = {Oct},
  publisher = {American Physical Society},
  doi = {10.1103/PhysRevX.10.041020},
  url = {https://link.aps.org/doi/10.1103/PhysRevX.10.041020}
}

@article{Jian_2022,
  title = {Criticality and entanglement in nonunitary quantum circuits and tensor networks of noninteracting fermions},
  author = {Jian, Chao-Ming and Bauer, Bela and Keselman, Anna and Ludwig, Andreas W. W.},
  journal = {Phys. Rev. B},
  volume = {106},
  issue = {13},
  pages = {134206},
  numpages = {38},
  year = {2022},
  month = {Oct},
  publisher = {American Physical Society},
  doi = {10.1103/PhysRevB.106.134206},
  url = {https://link.aps.org/doi/10.1103/PhysRevB.106.134206}
}

@article{Zirnbauer_1996,
    author = {Zirnbauer, Martin R.},
    title = {Riemannian symmetric superspaces and their origin in random‐matrix theory},
    journal = {Journal of Mathematical Physics},
    volume = {37},
    number = {10},
    pages = {4986-5018},
    year = {1996},
    month = {10},
    issn = {0022-2488},
    doi = {10.1063/1.531675},
    url = {https://doi.org/10.1063/1.531675},
}

@article{Altland_1997,
  title = {Nonstandard symmetry classes in mesoscopic normal-superconducting hybrid structures},
  author = {Altland, Alexander and Zirnbauer, Martin R.},
  journal = {Phys. Rev. B},
  volume = {55},
  issue = {2},
  pages = {1142--1161},
  numpages = {0},
  year = {1997},
  month = {Jan},
  publisher = {American Physical Society},
  doi = {10.1103/PhysRevB.55.1142},
  url = {https://link.aps.org/doi/10.1103/PhysRevB.55.1142}
}

@article{Wiseman_1993_2,
  title = {Interpretation of quantum jump and diffusion processes illustrated on the Bloch sphere},
  author = {Wiseman, H. M. and Milburn, G. J.},
  journal = {Phys. Rev. A},
  volume = {47},
  issue = {3},
  pages = {1652--1666},
  numpages = {0},
  year = {1993},
  month = {Mar},
  publisher = {American Physical Society},
  doi = {10.1103/PhysRevA.47.1652},
  url = {https://link.aps.org/doi/10.1103/PhysRevA.47.1652}
}

@article{Fidkowski_2021,
  doi = {10.22331/q-2021-01-17-382},
  url = {https://doi.org/10.22331/q-2021-01-17-382},
  title = {How {D}ynamical {Q}uantum {M}emories {F}orget},
  author = {Fidkowski, Lukasz and Haah, Jeongwan and Hastings, Matthew B.},
  journal = {{Quantum}},
  issn = {2521-327X},
  publisher = {{Verein zur F{\"{o}}rderung des Open Access Publizierens in den Quantenwissenschaften}},
  volume = {5},
  pages = {382},
  month = jan,
  year = {2021}
}

@article{Barun_2002,
    author = {Brun, Todd A.},
    title = {A simple model of quantum trajectories},
    journal = {American Journal of Physics},
    volume = {70},
    number = {7},
    pages = {719-737},
    year = {2002},
    month = {07},
    issn = {0002-9505},
    doi = {10.1119/1.1475328},
    url = {https://doi.org/10.1119/1.1475328}
}

@article{Turkeshi_2021,
  title = {Measurement-induced entanglement transitions in the quantum Ising chain: From infinite to zero clicks},
  author = {Turkeshi, Xhek and Biella, Alberto and Fazio, Rosario and Dalmonte, Marcello and Schir\'o, Marco},
  journal = {Phys. Rev. B},
  volume = {103},
  issue = {22},
  pages = {224210},
  numpages = {13},
  year = {2021},
  month = {Jun},
  publisher = {American Physical Society},
  doi = {10.1103/PhysRevB.103.224210},
  url = {https://link.aps.org/doi/10.1103/PhysRevB.103.224210}
}

@article{Coppola_2022,
  title = {Growth of entanglement entropy under local projective measurements},
  author = {Coppola, Michele and Tirrito, Emanuele and Karevski, Dragi and Collura, Mario},
  journal = {Phys. Rev. B},
  volume = {105},
  issue = {9},
  pages = {094303},
  numpages = {9},
  year = {2022},
  month = {Mar},
  publisher = {American Physical Society},
  doi = {10.1103/PhysRevB.105.094303},
  url = {https://link.aps.org/doi/10.1103/PhysRevB.105.094303}
}

@article{Braunstein_2005,
  title = {Quantum information with continuous variables},
  author = {Braunstein, Samuel L. and van Loock, Peter},
  journal = {Rev. Mod. Phys.},
  volume = {77},
  issue = {2},
  pages = {513--577},
  numpages = {0},
  year = {2005},
  month = {Jun},
  publisher = {American Physical Society},
  doi = {10.1103/RevModPhys.77.513},
  url = {https://link.aps.org/doi/10.1103/RevModPhys.77.513}
}

@article{Eisert_2010,
  title = {Colloquium: Area laws for the entanglement entropy},
  author = {Eisert, J. and Cramer, M. and Plenio, M. B.},
  journal = {Rev. Mod. Phys.},
  volume = {82},
  issue = {1},
  pages = {277--306},
  numpages = {0},
  year = {2010},
  month = {Feb},
  publisher = {American Physical Society},
  doi = {10.1103/RevModPhys.82.277},
  url = {https://link.aps.org/doi/10.1103/RevModPhys.82.277}
}

@article{Weedbrook_2012,
  title = {Gaussian quantum information},
  author = {Weedbrook, Christian and Pirandola, Stefano and Garc\'{\i}a-Patr\'on, Ra\'ul and Cerf, Nicolas J. and Ralph, Timothy C. and Shapiro, Jeffrey H. and Lloyd, Seth},
  journal = {Rev. Mod. Phys.},
  volume = {84},
  issue = {2},
  pages = {621--669},
  numpages = {0},
  year = {2012},
  month = {May},
  publisher = {American Physical Society},
  doi = {10.1103/RevModPhys.84.621},
  url = {https://link.aps.org/doi/10.1103/RevModPhys.84.621}
}

@article{Lvovsky_2009,
  title = {Continuous-variable optical quantum-state tomography},
  author = {Lvovsky, A. I. and Raymer, M. G.},
  journal = {Rev. Mod. Phys.},
  volume = {81},
  issue = {1},
  pages = {299--332},
  numpages = {0},
  year = {2009},
  month = {Mar},
  publisher = {American Physical Society},
  doi = {10.1103/RevModPhys.81.299},
  url = {https://link.aps.org/doi/10.1103/RevModPhys.81.299}
}

@article{Vaidman_1994,
  title = {Teleportation of quantum states},
  author = {Vaidman, Lev},
  journal = {Phys. Rev. A},
  volume = {49},
  issue = {2},
  pages = {1473--1476},
  numpages = {0},
  year = {1994},
  month = {Feb},
  publisher = {American Physical Society},
  doi = {10.1103/PhysRevA.49.1473},
  url = {https://link.aps.org/doi/10.1103/PhysRevA.49.1473}
}

@article{Braunstein_1998,
  title = {Teleportation of Continuous Quantum Variables},
  author = {Braunstein, Samuel L. and Kimble, H. J.},
  journal = {Phys. Rev. Lett.},
  volume = {80},
  issue = {4},
  pages = {869--872},
  numpages = {0},
  year = {1998},
  month = {Jan},
  publisher = {American Physical Society},
  doi = {10.1103/PhysRevLett.80.869},
  url = {https://link.aps.org/doi/10.1103/PhysRevLett.80.869}
}

@article{
Furusawa_1998,
author = {A. Furusawa  and J. L. Sørensen  and S. L. Braunstein  and C. A. Fuchs  and H. J. Kimble  and E. S. Polzik },
title = {Unconditional Quantum Teleportation},
journal = {Science},
volume = {282},
number = {5389},
pages = {706-709},
year = {1998},
doi = {10.1126/science.282.5389.706},
URL = {https://www.science.org/doi/abs/10.1126/science.282.5389.706},
eprint = {https://www.science.org/doi/pdf/10.1126/science.282.5389.706}}

@article{Gisin_2002,
  title = {Quantum cryptography},
  author = {Gisin, Nicolas and Ribordy, Gr\'egoire and Tittel, Wolfgang and Zbinden, Hugo},
  journal = {Rev. Mod. Phys.},
  volume = {74},
  issue = {1},
  pages = {145--195},
  numpages = {0},
  year = {2002},
  month = {Mar},
  publisher = {American Physical Society},
  doi = {10.1103/RevModPhys.74.145},
  url = {https://link.aps.org/doi/10.1103/RevModPhys.74.145}
}

@article{Scarani_2009,
  title = {The security of practical quantum key distribution},
  author = {Scarani, Valerio and Bechmann-Pasquinucci, Helle and Cerf, Nicolas J. and Du\ifmmode \check{s}\else \v{s}\fi{}ek, Miloslav and L\"utkenhaus, Norbert and Peev, Momtchil},
  journal = {Rev. Mod. Phys.},
  volume = {81},
  issue = {3},
  pages = {1301--1350},
  numpages = {0},
  year = {2009},
  month = {Sep},
  publisher = {American Physical Society},
  doi = {10.1103/RevModPhys.81.1301},
  url = {https://link.aps.org/doi/10.1103/RevModPhys.81.1301}
}

@article{Grosshans_2002,
  title = {Continuous Variable Quantum Cryptography Using Coherent States},
  author = {Grosshans, Fr\'ed\'eric and Grangier, Philippe},
  journal = {Phys. Rev. Lett.},
  volume = {88},
  issue = {5},
  pages = {057902},
  numpages = {4},
  year = {2002},
  month = {Jan},
  publisher = {American Physical Society},
  doi = {10.1103/PhysRevLett.88.057902},
  url = {https://link.aps.org/doi/10.1103/PhysRevLett.88.057902}
}

@article{Fava_2023,
  title = {Nonlinear Sigma Models for Monitored Dynamics of Free Fermions},
  author = {Fava, Michele and Piroli, Lorenzo and Swann, Tobias and Bernard, Denis and Nahum, Adam},
  journal = {Phys. Rev. X},
  volume = {13},
  issue = {4},
  pages = {041045},
  numpages = {33},
  year = {2023},
  month = {Dec},
  publisher = {American Physical Society},
  doi = {10.1103/PhysRevX.13.041045},
  url = {https://link.aps.org/doi/10.1103/PhysRevX.13.041045}
}

@article{Niederegger_2026,
  title = {Absence of measurement- and unraveling-induced entanglement transitions in continuously monitored one-dimensional free fermions},
  author = {Niederegger, Clemens and Vovk, Tatiana and Starchl, Elias and Sieberer, Lukas M.},
  journal = {Phys. Rev. B},
  volume = {113},
  issue = {14},
  pages = {144317},
  numpages = {24},
  year = {2026},
  month = {Apr},
  publisher = {American Physical Society},
  doi = {10.1103/636v-6s73},
  url = {https://link.aps.org/doi/10.1103/636v-6s73}
}

@article{Starchl_2025,
  title = {Generalized Zeno Effect and Entanglement Dynamics Induced by Fermion Counting},
  author = {Starchl, Elias and Fischer, Mark H. and Sieberer, Lukas M.},
  journal = {PRX Quantum},
  volume = {6},
  issue = {3},
  pages = {030302},
  numpages = {47},
  year = {2025},
  month = {Jul},
  publisher = {American Physical Society},
  doi = {10.1103/jppz-vdgn},
  url = {https://link.aps.org/doi/10.1103/jppz-vdgn}
}

@article{Peschel_2009,
doi = {10.1088/1751-8113/42/50/504003},
url = {https://doi.org/10.1088/1751-8113/42/50/504003},
year = {2009},
month = {dec},
publisher = {},
volume = {42},
number = {50},
pages = {504003},
author = {Peschel, Ingo and Eisler, Viktor},
title = {Reduced density matrices and entanglement entropy in free lattice models},
journal = {Journal of Physics A: Mathematical and Theoretical}
}

@article{Poboiko_2023,
  title = {Theory of Free Fermions under Random Projective Measurements},
  author = {Poboiko, Igor and P\"opperl, Paul and Gornyi, Igor V. and Mirlin, Alexander D.},
  journal = {Phys. Rev. X},
  volume = {13},
  issue = {4},
  pages = {041046},
  numpages = {26},
  year = {2023},
  month = {Dec},
  publisher = {American Physical Society},
  doi = {10.1103/PhysRevX.13.041046},
  url = {https://link.aps.org/doi/10.1103/PhysRevX.13.041046}
}

@article{Poboiko_2024,
  title = {Measurement-Induced Phase Transition for Free Fermions above One Dimension},
  author = {Poboiko, Igor and Gornyi, Igor V. and Mirlin, Alexander D.},
  journal = {Phys. Rev. Lett.},
  volume = {132},
  issue = {11},
  pages = {110403},
  numpages = {8},
  year = {2024},
  month = {Mar},
  publisher = {American Physical Society},
  doi = {10.1103/PhysRevLett.132.110403},
  url = {https://link.aps.org/doi/10.1103/PhysRevLett.132.110403}
}

@article{Poboiko_2025,
  title = {Measurement-induced transitions for interacting fermions},
  author = {Poboiko, Igor and P\"opperl, Paul and Gornyi, Igor V. and Mirlin, Alexander D.},
  journal = {Phys. Rev. B},
  volume = {111},
  issue = {2},
  pages = {024204},
  numpages = {39},
  year = {2025},
  month = {Jan},
  publisher = {American Physical Society},
  doi = {10.1103/PhysRevB.111.024204},
  url = {https://link.aps.org/doi/10.1103/PhysRevB.111.024204}
}

@misc{Jian_2023,
      title={Measurement-induced entanglement transitions in quantum circuits of non-interacting fermions: Born-rule versus forced measurements}, 
      author={Chao-Ming Jian and Hassan Shapourian and Bela Bauer and Andreas W. W. Ludwig},
      year={2023},
      eprint={2302.09094},
      archivePrefix={arXiv},
      primaryClass={cond-mat.stat-mech},
      url={https://arxiv.org/abs/2302.09094}, 
}

@book{Hanggi_1998,
	title={Quantum transport and dissipation},
	author={Dittrich, Thomas and H{\"a}nggi, Peter and Ingold, Gert-Ludwig and Kramer, Bernhard and Sch{\"o}n, Gerd and Zwerger, Wilhelm},
	volume={3},
	year={1998},
	publisher={Wiley-Vch Weinheim},
    address   = {Weinheim, Germany},
    isbn      = {9783527294354},
    doi={chrome-extension://oemmndcbldboiebfnladdacbdfmadadm/https://www.physik.uni-augsburg.de/theo1/hanggi/Papers/Chapter5.pdf}

}

@book{Stockmann_1999,
	place={Cambridge},
	title={Quantum Chaos: An Introduction}, publisher={Cambridge University Press}, author={Stöckmann, Hans-Jürgen},
	year={1999},
    address   = {Cambridge, UK},
    isbn      = {9780521592846},
    doi       = {10.1017/CBO9780511524622}

}

@article{Kiagawa_2011,
  title = {Transport properties of nonequilibrium systems under the application of light: Photoinduced quantum Hall insulators without Landau levels},
  author = {Kitagawa, Takuya and Oka, Takashi and Brataas, Arne and Fu, Liang and Demler, Eugene},
  journal = {Phys. Rev. B},
  volume = {84},
  issue = {23},
  pages = {235108},
  numpages = {13},
  year = {2011},
  month = {Dec},
  publisher = {American Physical Society},
  doi = {10.1103/PhysRevB.84.235108},
  url = {https://link.aps.org/doi/10.1103/PhysRevB.84.235108}
}

@article{Karen_2020,
	title = {Realization of an anomalous Floquet topological system with ultracold atoms},
	author = {Wintersperger, Karen and Braun, Christoph and Ünal, F. Nur and Eckardt, André and Liberto, Marco Di and Goldman, Nathan and Bloch, Immanuel and Aidelsburger, Monika},
	journal = {Nature Physics},
	volume = {16},
	issue = {10},
    pages = {1058},
    numpages = {6},
	year = {2020},
	publisher = {Nature Physics},
	doi = {10.1038/s41567-020-0949-y},
	url = {https://doi.org/10.1038/s41567-020-0949-y}
}

@article{Ivanov_2008,
	title = {Coherent Delocalization of Atomic Wave Packets in Driven Lattice Potentials},
	author = {Ivanov, V. V. and Alberti, A. and Schioppo, M. and Ferrari, G. and Artoni, M. and Chiofalo, M. L. and Tino, G. M.},
	journal = {Phys. Rev. Lett.},
	volume = {100},
	issue = {4},
	pages = {043602},
	numpages = {4},
	year = {2008},
	month = {Jan},
	publisher = {American Physical Society},
	doi = {10.1103/PhysRevLett.100.043602},
	url = {https://link.aps.org/doi/10.1103/PhysRevLett.100.043602}
}

@article{Poli_2011,
	title = {Precision Measurement of Gravity with Cold Atoms in an Optical Lattice and Comparison with a Classical Gravimeter},
	author = {Poli, N. and Wang, F.-Y. and Tarallo, M. G. and Alberti, A. and Prevedelli, M. and Tino, G. M.},
	journal = {Phys. Rev. Lett.},
	volume = {106},
	issue = {3},
	pages = {038501},
	numpages = {4},
	year = {2011},
	month = {Jan},
    publisher = {American Physical Society},
	doi = {10.1103/PhysRevLett.106.038501},
	url = {https://link.aps.org/doi/10.1103/PhysRevLett.106.038501}
}

@article{Chen_2011,
	title = {Controlling Correlated Tunneling and Superexchange Interactions with ac-Driven Optical Lattices},
	author = {Chen, Yu-Ao and Nascimb\`ene, Sylvain and Aidelsburger, Monika and Atala, Marcos and Trotzky, Stefan and Bloch, Immanuel},
	journal = {Phys. Rev. Lett.},
	volume = {107},
	issue = {21},
	pages = {210405},
	numpages = {5},
	year = {2011},
	month = {Nov},
	publisher = {American Physical Society},
	doi = {10.1103/PhysRevLett.107.210405},
	url = {https://link.aps.org/doi/10.1103/PhysRevLett.107.210405}
}

@article{Bitter_2016,
	title = {Experimental Observation of Dynamical Localization in Laser-Kicked Molecular Rotors},
	author = {Bitter, M. and Milner, V.},
	journal = {Phys. Rev. Lett.},
	volume = {117},
    issue = {14},
	pages = {144104},
	numpages = {5},
	year = {2016},
	month = {Sep},
	publisher = {American Physical Society},
	doi = {10.1103/PhysRevLett.117.144104},
	url = {https://link.aps.org/doi/10.1103/PhysRevLett.117.144104}
}

@article{Kundu_2014,
  title = {Effective Theory of Floquet Topological Transitions},
  author = {Kundu, Arijit and Fertig, H. A. and Seradjeh, Babak},
  journal = {Phys. Rev. Lett.},
  volume = {113},
  issue = {23},
  pages = {236803},
  numpages = {5},
  year = {2014},
  month = {Dec},
  publisher = {American Physical Society},
  doi = {10.1103/PhysRevLett.113.236803},
  url = {https://link.aps.org/doi/10.1103/PhysRevLett.113.236803}
}

@article{Scopa_2019,
  title = {Exact solution of time-dependent Lindblad equations with closed algebras},
  author = {Scopa, Stefano and Landi, Gabriel T. and Hammoumi, Adam and Karevski, Dragi},
  journal = {Phys. Rev. A},
  volume = {99},
  issue = {2},
  pages = {022105},
  numpages = {14},
  year = {2019},
  month = {Feb},
  publisher = {American Physical Society},
  doi = {10.1103/PhysRevA.99.022105},
  url = {https://link.aps.org/doi/10.1103/PhysRevA.99.022105}
}

@article{Ideka_2021,
	title = {Nonequilibrium steady states in the Floquet-Lindblad systems: van Vleck's high-frequency expansion approach},
	pages = {033},
	author = {Ikeda, Tatsuhiko and Chinzei, Koki and Sato, Masahiro},
	journal = {SciPost Phys. Core},
	volume = {4},
	year = {2021},
	publisher = {SciPost},
	doi = {10.21468/SciPostPhysCore.4.4.033},
	url = {https://scipost.org/10.21468/SciPostPhysCore.4.4.033}
}

@article{Magazzu_2018,
  title = {Asymptotic Floquet states of a periodically driven spin-boson system in the nonperturbative coupling regime},
  author = {Magazz\`u, Luca and Denisov, Sergey and H\"anggi, Peter},
  journal = {Phys. Rev. E},
  volume = {98},
  issue = {2},
  pages = {022111},
  numpages = {10},
  year = {2018},
  month = {Aug},
  publisher = {American Physical Society},
  doi = {10.1103/PhysRevE.98.022111},
  url = {https://link.aps.org/doi/10.1103/PhysRevE.98.022111}
}

@article{Iwahori_2016,
  title = {Long-time asymptotic state of periodically driven open quantum systems},
  author = {Iwahori, Koudai and Kawakami, Norio},
  journal = {Phys. Rev. B},
  volume = {94},
  issue = {18},
  pages = {184304},
  numpages = {6},
  year = {2016},
  month = {Nov},
  publisher = {American Physical Society},
  doi = {10.1103/PhysRevB.94.184304},
  url = {https://link.aps.org/doi/10.1103/PhysRevB.94.184304}
}

@article{Reimer_2018,
  title = {Nonadiabatic effects in periodically driven dissipative open quantum systems},
  author = {Reimer, Viktor and Pedersen, Kim G. L. and Tanger, Niklas and Pletyukhov, Mikhail and Gritsev, Vladimir},
  journal = {Phys. Rev. A},
  volume = {97},
  issue = {4},
  pages = {043851},
  numpages = {12},
  year = {2018},
  month = {Apr},
  publisher = {American Physical Society},
  doi = {10.1103/PhysRevA.97.043851},
  url = {https://link.aps.org/doi/10.1103/PhysRevA.97.043851}
}

@article{Prosen_2011,
  title = {Nonequilibrium Phase Transition in a Periodically Driven $XY$ Spin Chain},
  author = {Prosen, Tomaz and Ilievski, Enej},
  journal = {Phys. Rev. Lett.},
  volume = {107},
  issue = {6},
  pages = {060403},
  numpages = {4},
  year = {2011},
  month = {Aug},
  publisher = {American Physical Society},
  doi = {10.1103/PhysRevLett.107.060403},
  url = {https://link.aps.org/doi/10.1103/PhysRevLett.107.060403}
}

@article{Amico_2008,
  title = {Entanglement in many-body systems},
  author = {Amico, Luigi and Fazio, Rosario and Osterloh, Andreas and Vedral, Vlatko},
  journal = {Rev. Mod. Phys.},
  volume = {80},
  issue = {2},
  pages = {517--576},
  numpages = {0},
  year = {2008},
  month = {May},
  publisher = {American Physical Society},
  doi = {10.1103/RevModPhys.80.517},
  url = {https://link.aps.org/doi/10.1103/RevModPhys.80.517}
}

@article{Poboiko_2026,
  title = {Quantum dynamics of monitored free fermions: Evolution of quantum correlations and scaling at measurement-induced phase transitions},
  author = {Poboiko, Igor and Mirlin, Alexander D.},
  journal = {Phys. Rev. B},
  volume = {113},
  issue = {14},
  pages = {144311},
  numpages = {15},
  year = {2026},
  month = {Apr},
  publisher = {American Physical Society},
  doi = {10.1103/w1yq-xxbk},
  url = {https://link.aps.org/doi/10.1103/w1yq-xxbk}
}

@article{Carollo_2022,
  title = {Entangled multiplets and spreading of quantum correlations in a continuously monitored tight-binding chain},
  author = {Carollo, Federico and Alba, Vincenzo},
  journal = {Phys. Rev. B},
  volume = {106},
  issue = {22},
  pages = {L220304},
  numpages = {7},
  year = {2022},
  month = {Dec},
  publisher = {American Physical Society},
  doi = {10.1103/PhysRevB.106.L220304},
  url = {https://link.aps.org/doi/10.1103/PhysRevB.106.L220304}
}

@article{Chahine_2023,
  title = {Entanglement phases, localization, and multifractality of monitored free fermions in two dimensions},
  author = {Chahine, K. and Buchhold, M.},
  journal = {Phys. Rev. B},
  volume = {110},
  issue = {5},
  pages = {054313},
  numpages = {10},
  year = {2024},
  month = {Aug},
  publisher = {American Physical Society},
  doi = {10.1103/PhysRevB.110.054313},
  url = {https://link.aps.org/doi/10.1103/PhysRevB.110.054313}
}

@article{Lumia_2023,
  title = {Measurement-induced transitions beyond {Gaussianity}: {A} single particle description},
  author = {Lumia, Luca and Tirrito, Emanuele and Fazio, Rosario and Collura, Mario},
  journal = {Phys. Rev. Res.},
  volume = {6},
  issue = {2},
  pages = {023176},
  numpages = {10},
  year = {2024},
  month = may,
  publisher = {American Physical Society},
  doi = {10.1103/PhysRevResearch.6.023176},
  url = {https://link.aps.org/doi/10.1103/PhysRevResearch.6.023176}
}

@article{Guo_2025,
  title = {Field theory of monitored interacting fermion dynamics with charge conservation},
  author = {Guo, Haoyu and Foster, Matthew S. and Jian, Chao-Ming and Ludwig, Andreas W. W.},
  journal = {Phys. Rev. B},
  volume = {112},
  issue = {6},
  pages = {064304},
  numpages = {31},
  year = {2025},
  month = {Aug},
  publisher = {American Physical Society},
  doi = {10.1103/gdxd-pw8v},
  url = {https://link.aps.org/doi/10.1103/gdxd-pw8v}
}

@article{KamenevLevchenko,
    author      = { Kamenev, Alex and Levchenko, Alex},
    title       = { Keldysh technique and non-linear $\sigma$-model: basic principles and applications},
    journal     = {Advances in Physics},
    volume      = {58},
    number      = {3},
    pages       = {197-319},
    year        = {2009},
    publisher   = {Taylor & Francis},
    doi         = {10.1080/00018730902850504},
}

@article{KlichLevitov,
    title       = {Quantum Noise as an Entanglement Meter},
    author      = {Klich, Israel and Levitov, Leonid},
    journal     = {Phys. Rev. Lett.},
    volume      = {102},
    issue       = {10},
    pages       = {100502},
    numpages    = {4},
    year        = {2009},
    month       = {Mar},
    publisher   = {American Physical Society},
    doi         = {10.1103/PhysRevLett.102.100502},
    url         = {https://link.aps.org/doi/10.1103/PhysRevLett.102.100502}
}

@article{FavaNahum2024,
  title = {{Monitored fermions with conserved $\mathrm{U}(1)$ charge}},
  author = {Fava, Michele and Piroli, Lorenzo and Bernard, Denis and Nahum, Adam},
  journal = {Phys. Rev. Res.},
  volume = {6},
  issue = {4},
  pages = {043246},
  numpages = {21},
  year = {2024},
  month = {Dec},
  publisher = {American Physical Society},
  doi = {10.1103/PhysRevResearch.6.043246},
  url = {https://link.aps.org/doi/10.1103/PhysRevResearch.6.043246}
}

\end{document}